\documentclass[review]{elsarticle}

\usepackage{lineno,hyperref}
\modulolinenumbers[5]

\journal{Optics Communications}

\usepackage{amsfonts}
\usepackage{amsmath}
\usepackage{amssymb}
\usepackage{booktabs}
\usepackage{tabularx}
\usepackage{mathtools}
\usepackage{pgfplots}
\usepackage{geometry}
\usepackage{pdflscape}
\usepackage{enumerate}
\usepackage{esvect}
\usepackage[figuresright]{rotating}
\usepackage{footnote}
\usepackage{listings}
\usepackage{xcolor}
\usepackage{enumitem}
\usepackage{tocloft}
\usepackage{textcomp}
\usepackage{braket}
\usepackage{nicefrac}
\usepackage{nccmath}
\usepackage{bm}
\usepackage{import}
\usepackage{dpfloat}
\usepackage{nicefrac}
\usepackage{subcaption}
\usepackage{float}
\usepackage{afterpage}
\usepackage{multirow}
\usepackage{siunitx}
\usepackage{graphicx}
\usepackage{tikz}
\usepackage[ruled]{algorithm2e}
\graphicspath{{Figs/}}
\DeclarePairedDelimiter\abs{\lvert}{\rvert}%
\allowdisplaybreaks

\begin{document}
    
    \begin{frontmatter}
        
        \title{Holographic Predictive Search: Extending the Scope}
                
        \author[mymainaddress]{Peter J. Christopher\corref{mycorrespondingauthor}}
        \cortext[mycorrespondingauthor]{Corresponding author}
        \ead{pjc209@cam.ac.uk}
        \ead[url]{www.peterjchristopher.me.uk}
        
        \author[mymainaddress]{Ralf Mouthaan}
        
        \author[mymainaddres2]{George S. D. Gordon}
        
        \author[mymainaddress]{Timothy D. Wilkinson}
        
        \address[mymainaddress]{Centre of Molecular Materials, Photonics and Electronics, University of Cambridge}
        \address[mymainaddres2]{Department of Electrical and Electronic Engineering, University of Nottingham}
        
        \begin{abstract}
            Holographic Predictive Search (HPS) is a novel approach to search-based hologram generation that uses a mathematical understanding of the optical transforms to make informed optimisation decisions. Existing search techniques such as Direct Search (DS) and Simulated Annealing (SA) rely on trialling modifications to a test hologram and observing the results. A formula is used to decide whether the change should be accepted. HPS operates presciently, using knowledge of the underlying mathematical relationship to make exact changes to the test hologram that guarantee the 'best' outcome for that change. 
            
            In this work, we extend the scope of the original research to cover both phase and amplitude modulating Spatial Light Modulators (SLMs), both phase sensitive and phase insensitive systems and both Fresnel and Fraunhofer diffraction.  In the cases discussed, improvements of up to 10x are observed in final error and the approach also offers significant performance benefits in generation time. This comes at the expense of increased complexity and loss of generality.
        \end{abstract}
    
    
    \end{frontmatter}

    \section{Introduction}
    
    The expansion of Computer Generated Holography (CGH) in recent years has seen application in areas including super resolution microscopy, optical tweezing, quantum mechanics and optical communication \cite{milione2015using, backlund2016removing, cheng2017vortex,wang2016advances,wang2018optically,xie2014harnessing}. For applications where quality is the primary consideration, Holographic Search Algorithms (HSAs) are a common approach with algorithms like such as Direct Search (DS) and Simulated Annealing (SA) being used~\cite{DirectSearch_2,kirkpatrick1983optimization}.
    
    In our recent paper we introduced Holographic Predictive Search (HPS), an algorithm that offers the potential to improve upon existing HSAs \cite{HPS1}. That work exclusively considered the case of phase modulated holograms where the replay field phase is of interest. Here we expand on this to treat both phase and amplitude modulated holograms; both phase sensitive and phase insensitive replay fields and both the Fresnel and Fraunhofer diffraction regimes.
    
     HPS uses a prescient model of the Fourier and Fresnel Transforms used in far- and mid- field holography to provide a predictive alternative to traditional \textit{blind} search approaches. While HPS offers considerable performance improvements over rival HSAs  of up to $10\times$ lower convergence times it comes at the expense of reduced flexibility. Here we expand on the single case presented initially to provide algorithmic variants for an array of system combinations. Each is presented with an analysis of performance and of the relative advantages of HPS over other HSAs. Reviews of CGH are available \cite{doi:10.1080/15980316.2016.1255672, Tsang:18} so we start with only the bare minimum of background required.
        
    \section{Background}
    
    The Discrete Fourier Transform (DFT) forms the core of the holographic process,
    
    \begin{align}
        F_{u,v} & = \frac{1}{\sqrt{N_xN_y}}\sum_{x=0}^{N_x-1}\sum_{y=0}^{N_y-1} f_{xy}e^{-2\pi i \left(\frac{u x}{N_x} + \frac{v y}{N_y}\right)} \label{fouriertrans2d5c}   \\
        f_{x,y} & = \frac{1}{\sqrt{N_xN_y}}\sum_{u=0}^{N_x-1}\sum_{v=0}^{N_y-1} F_{uv}e^{2\pi i \left(\frac{u x}{N_x} + \frac{v y}{N_y}\right)}  \label{fouriertrans2d5d}
    \end{align}

    where $u$ and $v$ represent the spatial frequencies and $x$ and $y$ represent the source coordinates. Fast Fourier Transforms (FFTs) are typically used to calculate the DFT with calculation times of $O(N_xN_y\log{N_xN_y})$ where $N_x$ and $N_y$ are the respective $x$ and $y$ resolutions~\cite{carpenter2010graphics,frigo2005design}.
    
    The far-field pattern produced by passing coherent light through a Spatial Light Modulator (SLM) is equivalent to taking the DFT of the SLM aperture function multiplied by the static pixel shape parameter and coherent illumination~\cite{goodman2005introduction}. For an ideal pixellated SLM acting on uniform unit intensity planar wavefronts with $100\%$ fill factor pixels, the produced hologram is given by the DFT of the SLM aperture function. More generally, the projected hologram is often referred to as the \textit{Replay Field} or \textit{Replay Plane} and the SLM as the \textit{Diffraction Field} or \textit{Diffraction Plane}.
    
    Real-world SLMs modulate light only in a limited fashion, typically phase or amplitude only~\cite{Huang2018,deBougrenetdelaTocnaye:97}. When addressed digitally, this is restricted further to discrete levels. Finding an SLM aperture function for a given far-field hologram $F(u,v)$ is identical to the problem of finding $f(x,y)$ where $F(u,v) = \mathcal{F}\{f(x,y)\}$ subject to these constraints. $\mathcal{F}$ here refers to the Fourier transform.
    
    \section{Holographic Search Algorithms}
    
\begin{figure}[htbp]
        \centering
        {\includegraphics[trim={0 0 0 0},width=0.7\linewidth,page=1]{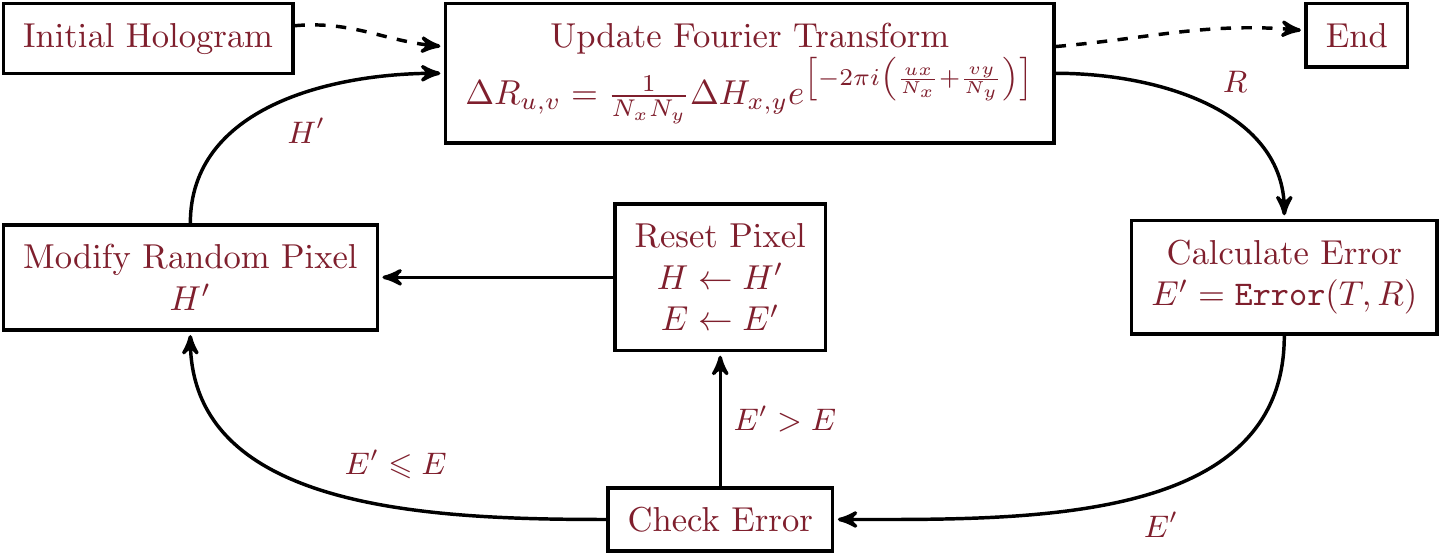}}
        \caption{Fast Direct Search}
        \label{fig:dsfast}
    \end{figure}

\begin{figure}[htbp]
        \centering
        {\includegraphics[trim={0 0 0 0},width=0.7\linewidth,page=1]{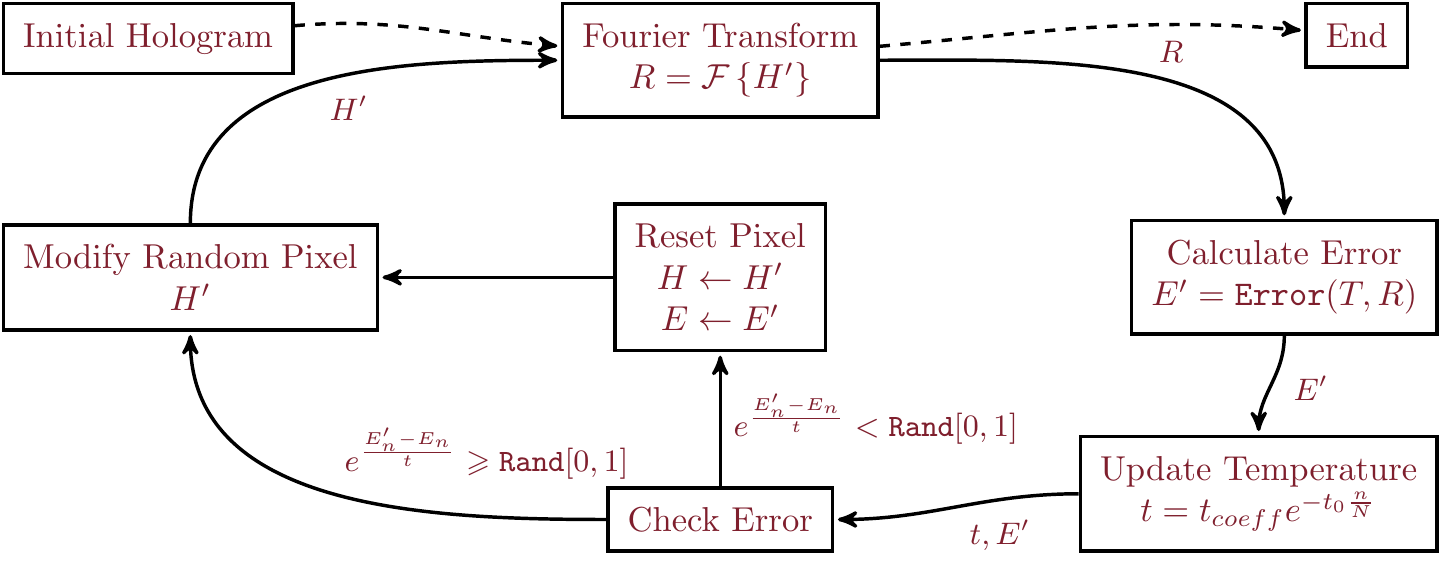}}
        \caption{Simulated Annealing}
        \label{fig:safig}
    \end{figure}

    In our recent paper we introduced Holographic Predictive Search (HPS) and compared it with DS and SA. The procedure for these two algorithms is shown in Figures~\ref{fig:dsfast} and \ref{fig:safig}. Key to these algorithms is the fact that we can avoid performing a full DFT at each iteration, instead using the following $O(N_xN_y)$ update step
    
    \begin{equation} \label{updatestep}
        \Delta R_{u,v} = \frac{1}{\sqrt{N_xN_y}}\Delta H_{x,y} e^{\left[-2\pi i\left(\frac{ux}{N_x}+\frac{vy}{N_y}\right)\right]}
    \end{equation}
    
    where change $\Delta H_{x,y}$ in aperture function causes a change $\Delta R_{u,v}$ in the replay field. 
    
    One fundamental feature of these two algorithms is that they are blind, randomly selecting a new pixel value to trial. The change in error or quality metric is then observed and the change is either discarded or accepted. In the case of DS, the decision is binary with change always being accepted if the error is reduced. SA adds a probabalistic element that can sometimes select worse solutions. This slows the algorithm but reduces the chance of being captured in local minima.         
        
    \section{Holographic Predictive Search}
    
    Our earlier work introduced Holographic Predictive Search (HPS) for hologram generation \cite{HPS1}. HPS operates in a similar manner to DS and SA but instead of blindly choosing a new pixel value and observing the change in error, HPS operates by generating a linear relationship for the \textit{best} change in a pixel value. This led to significant performance improvements in convergence time and error reduction. The downside was an increase in computational complexity per iteration and a loss of generality in mathematical form. 
    
    As originally presented, this algorithm was only applicable in the case of a phase-modulated SLM, with a Fraunhofer hologram where the target image was phase sensitive.  Here we extend the scope of this approach to cover both phase and amplitude modulating SLMs, both phase sensitive and phase insensitive systems and both Fresnel and Fraunhofer diffraction.
    
    \begin{figure}[htbp]
        \centering
        {\includegraphics[trim={0 0 0 0},width=0.7\linewidth,page=1]{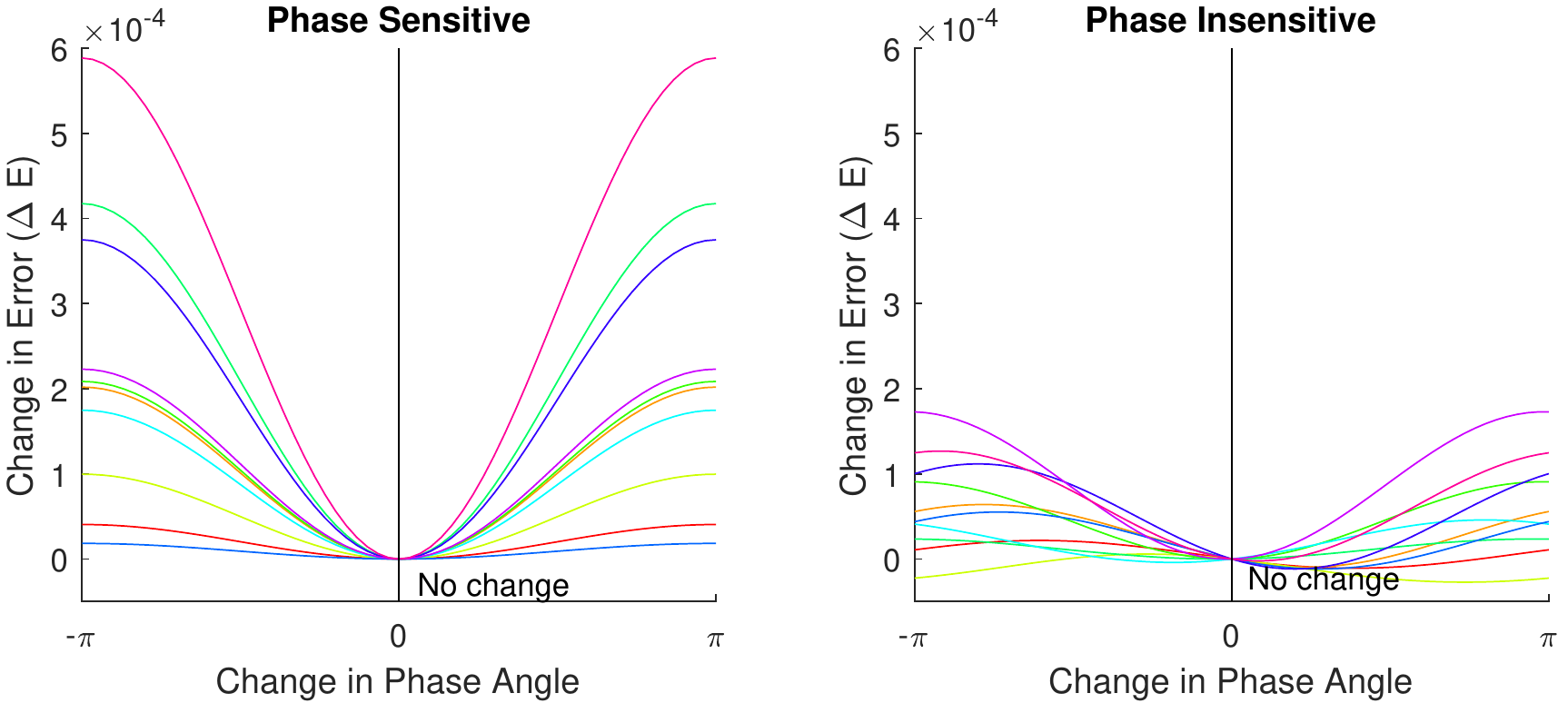}}
        \caption{Selection of final errors depending on pixel changes for phase modulating SLMs for phase sensitive replay fields (left) and phase insensitive replay fields (right). The SLM is assumed to be $256\times 256$ pixels with a flat unit intensity illumination.}
        \label{fig:pixelChange_002}
    \end{figure}

\begin{figure}[htbp]
        \centering
        {\includegraphics[trim={0 0 0 0},width=0.7\linewidth,page=1]{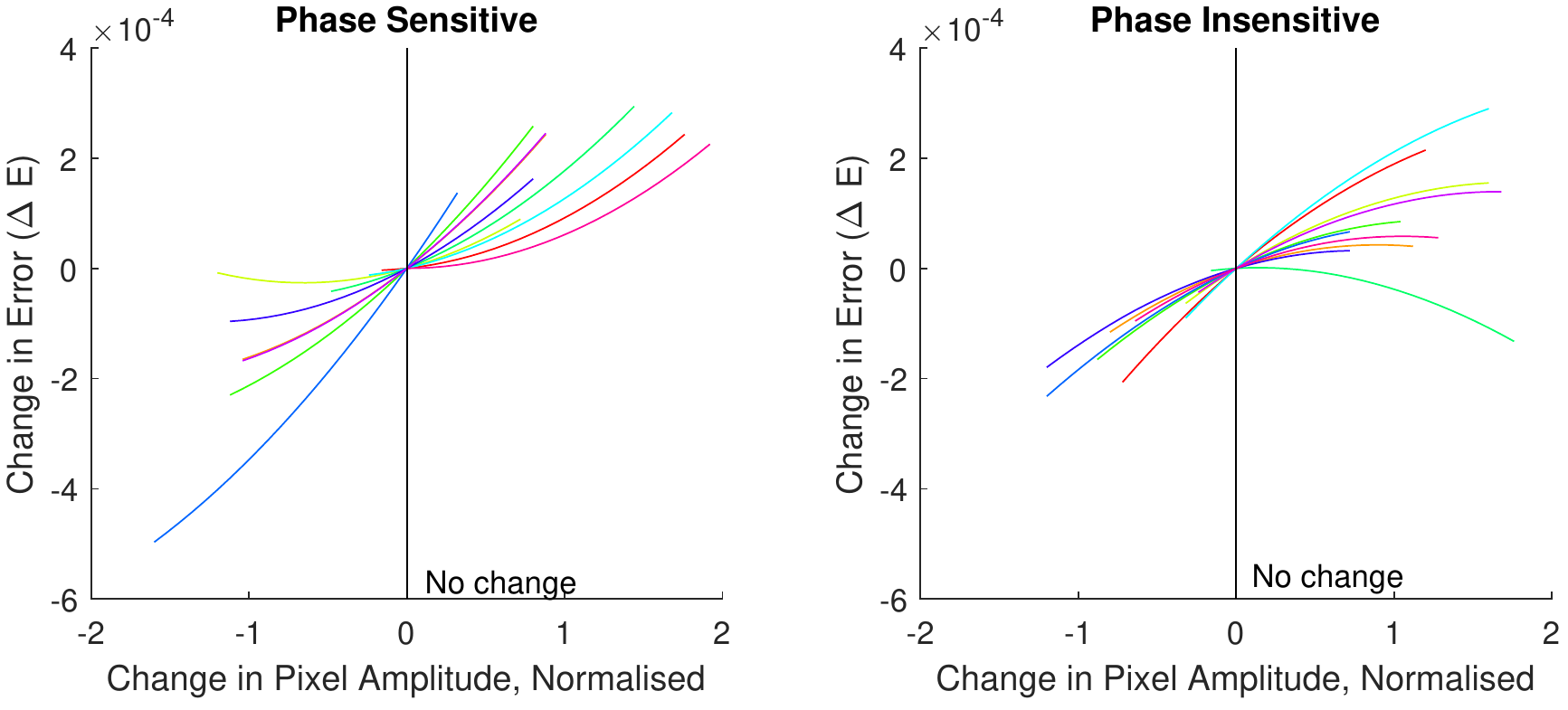}}
        \caption{Selection of final errors depending on pixel changes for amplitude modulating SLMs for phase sensitive replay fields (left) and phase insensitive replay fields (right). The SLM is assumed to be $256\times 256$ pixels with a flat unit intensity illumination.}
        \label{fig:pixelChange_001}
    \end{figure}

    To give an understanding of the differences between these two cases Figures \ref{fig:pixelChange_002} and \ref{fig:pixelChange_001} take an initial inverse Fourier transform of the \textit{Mandrill} test image and plot the effect on mean squared error (MSE) of changing the level on 10 randomly selected SLM pixels for two different categories of SLM - phase and amplitude modulating - and for two different categories of replay field - phase sensitive and insensitive. These were generated by selecting a random hologram pixel $H_{x,y}$ and plotting the change in final error for a range of pixel values. They show that the response to level changes of a single phase pixel has a near sinusoidal effect on the error of the replay field whereas changing an amplitude pixel has a more linear response. 
    
    The aim of HPS is to mathematically capture the nature of the curves shown and solve for the optimum location without having to apply Eq.~\ref{updatestep} to every step. In practice the phase insensitive cases degenerate to quartic polynomials and it is only through the judicious use of simplifications that we can provide high speed solutions.
        
    \subsection{Extending the Scope}
        
    When originally presented, only the phase modulating SLM with phase sensitive replay field was considered as that case simultaneously provided the simplest analysis and an array of potential applications. Our aim in this paper is to develop the relationships for the other paradigms and to present analysis of their efficacy. There are three sets of variants that we must consider.
        
    \begin{itemize}
        \item \textbf{Transform Type} - Only far-field or Fraunhofer holograms can be modelled as an Fourier Transform. For mid-field holograms, an additional quadratic phase term must be added to form a Fresnel Transform. 
        \item \textbf{SLM Modulation Behaviour} - SLMs typically modulate in either phase or amplitude. Each paradigm involves a different set of relationships between the SLM and the replay field.
        \item \textbf{Phase Sensitivity} - As originally presented, HPS aims to satisfy both the amplitude and phase constraints of the target. Many display applications, however, do not require the phase constraint due to the human eye's phase insensitivity. This additional freedom requires a separate formulation.
    \end{itemize}
        
    This paper sets out the necessary background to conform to every combinations of these constraints.
        
    \subsection{Error and Quality Metrics}
    
    The choice of whether phase sensitivity is considered also changes the error metric used where $T$ and $R$ are the target and actual replay fields
    
    \begin{align} \label{mse}
    E_{MSE,ps}(T,R) &= \frac{1}{N_x N_y}\sum_{u=0}^{N_x-1}\sum_{v=0}^{N_y-1} \left[\abs{T_{u,v} -  R_{u,v}}\right]^2 \nonumber \\
    E_{MSE,pi}(T,R) &= \frac{1}{N_x N_y}\sum_{u=0}^{N_x-1}\sum_{v=0}^{N_y-1} \left[\abs{T_{u,v}} -  \abs{R_{u,v}}\right]^2
    \end{align}
    
    Where $E_{MSE,ps}$ and $E_{MSE,pi}$ represent the \textit{phase sensitive} and the \textit{phase insensitive} mean squared errors (MSE) respectively.
    
    The Structural Similarity Index (SSIM) is often used in preference to MSE when image quality rather than numerical error is the primary concern \cite{wang2004image}. The algorithms presented here primarily target MSE and we return to this issue later. 
    
    \section{Methods}
    
\begin{figure}[htbp]
        \centering
        {\includegraphics[trim={0 0 0 0},width=\linewidth,page=1]{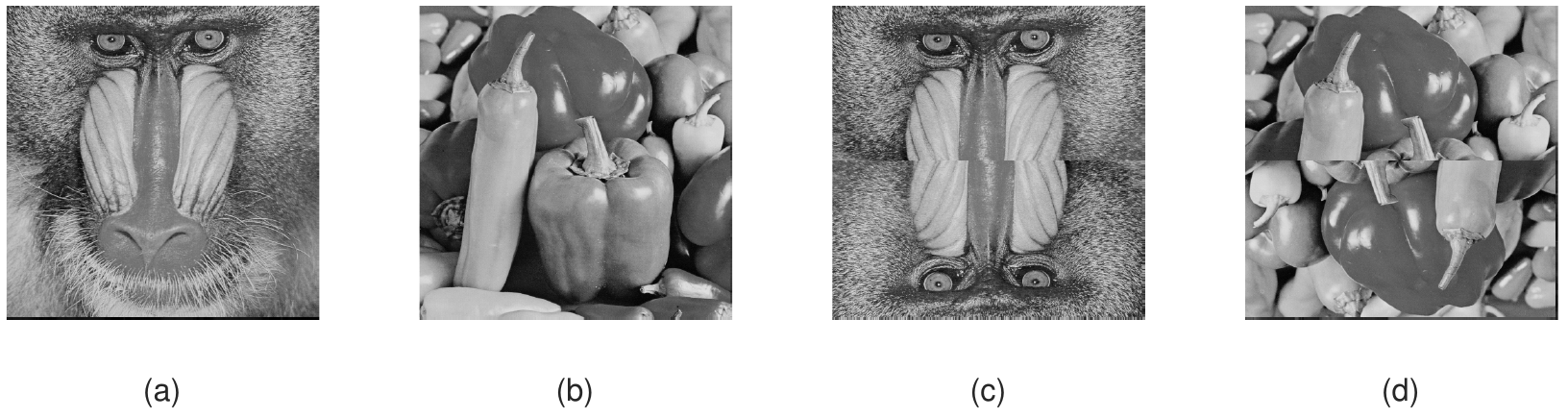}}
        \caption{The two test images used showing the \textit{Mandrill} and \textit{Peppers} as well as their artificially symmetric counterparts.}
        \label{fig:converge0}
    \end{figure}

    It is challenging to fairly compare the performance of techniques across different system designs. In order to best serve our readers we have adopted the following conventions:
    
    \begin{enumerate}
        \item We have used the \textit{Mandrill} test image shown in Figure~\ref{fig:converge0}a to provide the target intensities.
        \item Amplitude holograms are generated with a rotationally symmetric version of Figure~\ref{fig:converge0}c to avoid error due to image symmetry.
        \item Phase sensitive holograms are generated with the \textit{Peppers} test image used used as the phase component as shown in Figure~\ref{fig:converge0}b with rotationally symmetric variant shown in Figure~\ref{fig:converge0}d.
        \item For phase insensitive holograms, the entire target region is solved for. In the case of phase sensitive holograms, for reasons of degrees of freedom, we have scaled the target image to only fill the central quadrant of the initial replay field and set the surrounding regions to zero.
        \item Planar unit intensity incident on the hologram is assumed with the target scaled to ensure conservation of energy.
    \end{enumerate}
    
    These differences in method mean that the normalised error metrics used should be treated as distinct in each case and cannot be compared quantitatively between cases.. 
    
    \section{Fraunhofer Domain}
    
    \subsection{Phase Modulated SLM, Phase Sensitive Replay Field} \label{ps}
    
\begin{figure}[htbp]
        \centering
        {\includegraphics[trim={0 0 0 0},width=0.7\linewidth,page=1]{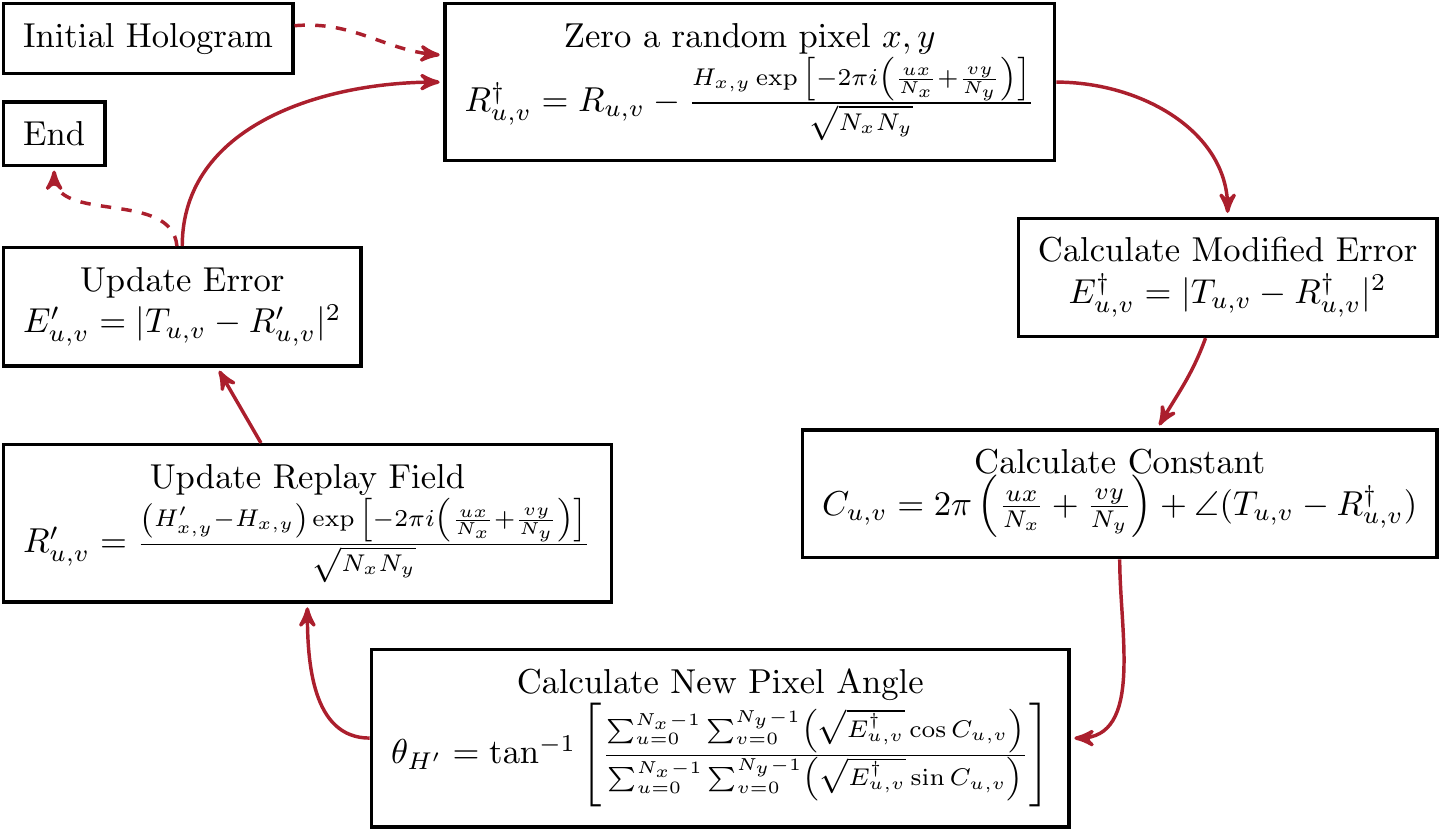}}
        \caption{Holographic Predictive Search for phase modulated, far-field holography with phase sensitive target. Modified with permission from \cite{HPS1}.}
        \label{fig:alg_HPS_ps}
    \end{figure}

\begin{figure}[tb]
        \centering
        {\includegraphics[trim={0 0 0 0},width=0.7\linewidth,page=1]{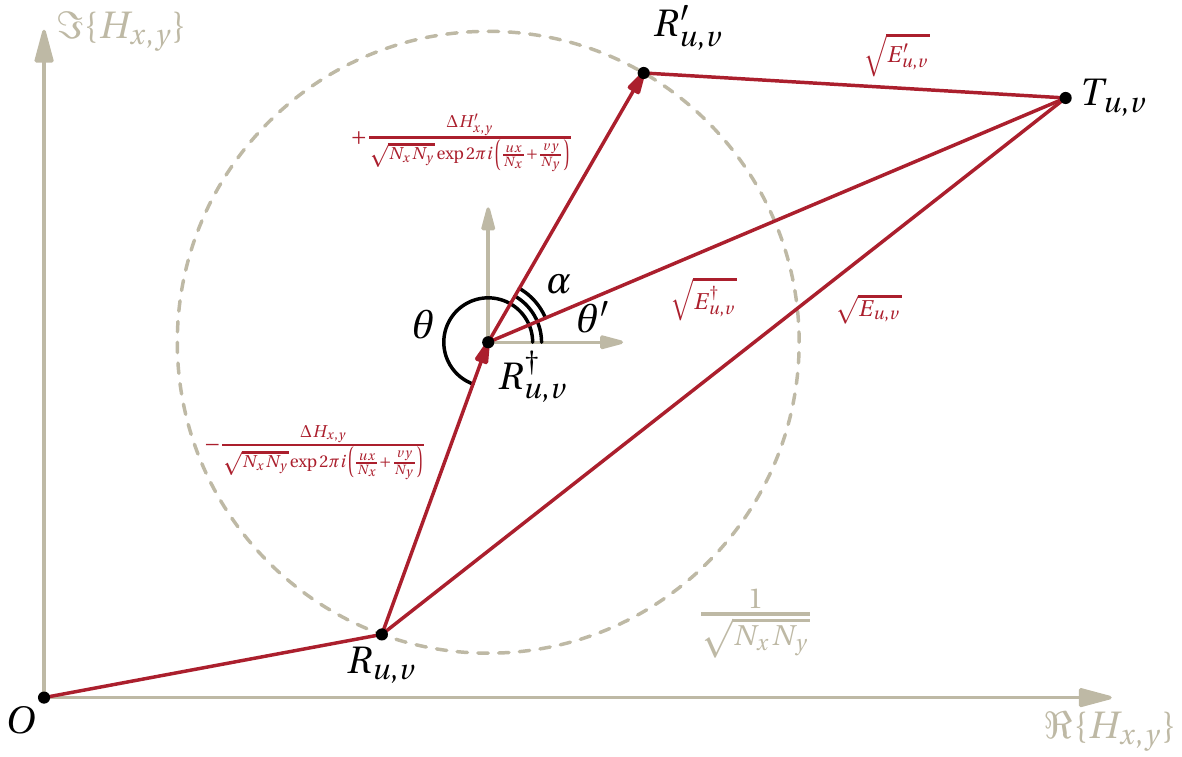}}
        \caption{Problem geometry in phase modulated, phase insensitive case. Used with permission from \cite{HPS1}.}
        \label{fig:predicted1}
    \end{figure}

    HPS was originally developed for the case of a Fraunhofer or far-field system displayed on a phase modulating SLM with a phase sensitive target \cite{HPS1}. That algorithm is shown in Figure~\ref{fig:alg_HPS_ps} with problem geometry as shown in Figure~\ref{fig:predicted1}.
    
    The single pixel $x,y$ on the SLM is set to zero and the resulting change in the replay field $R^{\dagger}$ and error $E^{\dagger}$ calculated using Eq.~\ref{updatestep}. An expression for the error after assigning a new phase to the selected pixel is then given by
    
    \begin{align} \label{bigsum}
        \Delta E' & = 1 
        - \frac{2}{\sqrt{N_xN_y}} 
        \Bigg[\cos{\theta_{H'}}\sum^{N_x-1}_{u=0}\sum^{N_y-1}_{v=0}\sqrt{E^{\dagger}_{u,v}}\cos{C_{u,v}}
        +\sin{\theta_{H'}}\sum^{N_x-1}_{u=0}\sum^{N_y-1}_{v=0}\sqrt{E^{\dagger}_{u,v}}\sin{C_{u,v}} \Bigg]
         \nonumber \\
        \quad C_{u,v} &= 2\pi\left(\frac{ux}{N_x}+\frac{vy}{N_y}\right) + \angle(T_{u,v}-R^{\dagger}_{u,v})
    \end{align}
    
    This formulation only uses terms known at runtime and allows for solutions to find the most desirable value of new SLM pixel phase $\theta_{H'}$. Note that $\angle{X}$ here refers to the phase angle of $X$.
    
    \begin{equation} 
    \theta_{H'}=\tan^{-1}{\left[\frac{\sum^{N_x-1}_{u=0}\sum^{N_y-1}_{v=0}\left(\sqrt{E^{\dagger}_{u,v}}\sin{C_{u,v}}\right)}{\sum^{N_x-1}_{u=0}\sum^{N_y-1}_{v=0}\left(\sqrt{E^{\dagger}_{u,v}}\cos{C_{u,v}}\right)}\right]}
    \end{equation}
    
    This allowed for greatly improved speed of convergence at the expense of additional iteration complexity.
    
    Below we discuss the seven other cases.
    
    \subsection{Phase Modulated SLM, Phase Insensitive Replay Field} \label{pi}
    
\begin{figure}[tb]
        \centering
        {\includegraphics[trim={0 0 0 0},width=0.7\linewidth,page=1]{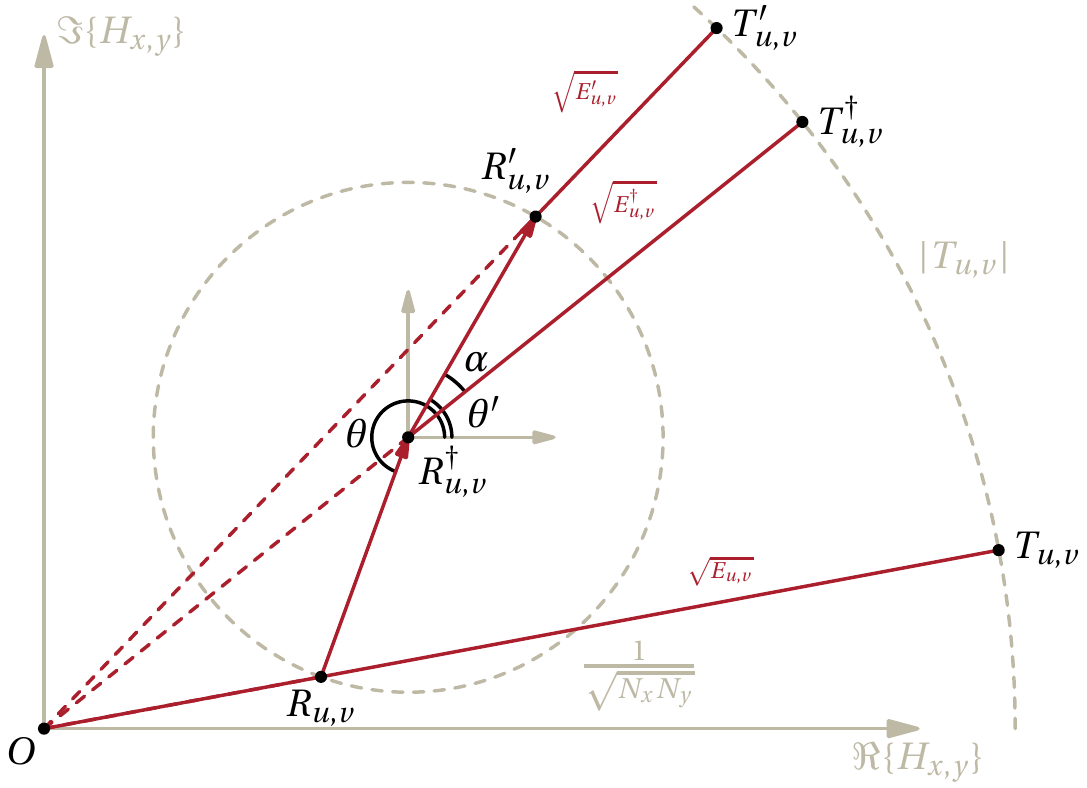}}
        \caption{Problem geometry in phase modulated, phase insensitive case.}
        \label{fig:predicted2}
    \end{figure}

    The phase insensitivity of the eye means that display applications are often phase insensitive. This greatly increases the problem freedom but also changes the predictive geometry into a non-linear problem. The updated regime is shown on the Argand diagram in Figure~\ref{fig:predicted2}.
    
    Zeroing an individual SLM pixel $x,y$ introduces error to location $u,v$ in the replay field $R$ given by Eq.~\ref{updatestep} with $\Delta H_{x,y}=-H_{x,y}$. This replay field we term $R^{\dagger}$. The task is to find new pixel value $H'_{x,y}$ of unit magnitude so that the error in the new replay field $R'$ is minimised. Expressing $\theta$ and $\theta'$, the old and new pixel phases, in terms of unknown $\angle{H'_{x,y}}$ and known $x$, $y$, $u$, $v$, $N_x$ and $N_y$,
    
    \begin{align} \label{XXXXX}
    \theta & = \angle{H_{x,y}}-2\pi\left(\frac{ux}{N_x}+\frac{vy}{N_y}\right), \nonumber \\
    \theta' &= \angle{H'_{x,y}}-2\pi\left(\frac{ux}{N_x}+\frac{vy}{N_y}\right)
    \end{align}

    The error after zeroing pixel $x,y$ is given as $E^{\dagger}_{u,v}=\left(\abs{T_{u,v}}-\abs{R^{\dagger}_{u,v}}\right)^2$. The new error $E'_{u,v}$ is given by
    
    \begin{align} \label{eqngdsafgsd}
    E'_{u,v}  =& \left[\abs{T_{u,v}}-\abs{R'_{u,v}}\right]^2 \nonumber \\
    =& \left[\abs{T_{u,v}}-\sqrt{\abs{R^{\dagger}_{u,v}}^2+\frac{1}{N_xN_y}+\frac{2\abs{R^{\dagger}_{u,v}}\cos{\alpha}}{\sqrt{N_xN_y}}}\right]^2 \nonumber \\
    \Rightarrow \Delta E'_{u,v} & =  E'_{u,v} - E^{\dagger}_{u,v} \nonumber \\
    =& \frac{1}{N_xN_y}+\frac{2\abs{R^{\dagger}_{u,v}}\cos{\alpha}}{\sqrt{N_xN_y}}+2\abs{T_{u,v}}\abs{R^{\dagger}_{u,v}}\nonumber \\
    &-2\abs{T_{u,v}}\sqrt{\abs{R^{\dagger}_{u,v}}^2+\frac{1}{N_xN_y}+\frac{2\abs{R^{\dagger}_{u,v}}\cos{\alpha}}{\sqrt{N_xN_y}}}
    \end{align}
    
    As 
    
    \begin{equation}
    \alpha = \theta' - \angle{R^{\dagger}_{u,v}} = \angle{H'_{x,y}}-\left[2\pi\left(\frac{ux}{N_x}+\frac{vy}{N_y}\right) + \angle{R^{\dagger}_{u,v}}\right]
    \end{equation}
    
    we can apply the Taylor expansion of $\sqrt{1 + z}$ to give
    
    \begin{equation} 
    \Delta E'_{u,v}  = D_{u,v}
    + F_{u,v}\left[\cos{\theta_{H'}}\cos{C_{u,v}}+\sin{\theta_{H'}}\sin{C_{u,v}}\right] \nonumber
    \end{equation}
    
    where
    
    \begin{align} \label{sdhsdfghdfgh}
    D_{u,v} &= \frac{1}{N_xN_y}+2\abs{T_{u,v}}\abs{R^{\dagger}_{u,v}}-2\abs{T_{u,v}}\sqrt{\abs{R^{\dagger}_{u,v}}^2+\frac{1}{N_xN_y}}  \nonumber \\
    F_{u,v} &=\frac{2\abs{R^{\dagger}_{u,v}}}{\sqrt{N_xN_y}}-\frac{2\abs{T_{u,v}}\abs{R^{\dagger}_{u,v}}}{\sqrt{N_xN_y}\sqrt{\abs{R^{\dagger}_{u,v}}^2+\frac{1}{N_xN_y}}}  \nonumber \\
    \quad C_{u,v} &= 2\pi\left(\frac{ux}{N_x}+\frac{vy}{N_y}\right) + \angle{R^{\dagger}_{u,v}}
    \end{align}
    
    Summing  over all pixels
    
    \begin{align}
    \Delta E' &= \sum^{N_x-1}_{u=0}\sum^{N_y-1}_{v=0} \Delta E'_{u,v} \nonumber \\
    & = \sum^{N_x-1}_{u=0}\sum^{N_y-1}_{v=0}D_{u,v}
    +\cos{\theta_{H'}}\sum^{N_x-1}_{u=0}\sum^{N_y-1}_{v=0}F_{u,v}\cos{C_{u,v}} 
    +\sin{\theta_{H'}}\sum^{N_x-1}_{u=0}\sum^{N_y-1}_{v=0}F_{u,v}\sin{C_{u,v}}
    \end{align}
    
    Taking $\nicefrac{\mathrm{d}\Delta E'}{\mathrm{d}\theta_{H'}}=0$ to find the the value of $\theta_{H'}$ where $\Delta E'$ is minimum
    
    \begin{equation} 
    \sin{\theta_{H'}}\sum^{N_x-1}_{u=0}\sum^{N_y-1}_{v=0}F_{u,v}\cos{C_{u,v}} - \cos{\theta_{H'}}\sum^{N_x-1}_{u=0}\sum^{N_y-1}_{v=0}F_{u,v}\sin{C_{u,v}} = 0
    \end{equation}
    
    which is trivially solvable.
    
    \begin{equation} 
    \theta_{H'}=\tan^{-1}{\left[\frac{\sum^{N_x-1}_{u=0}\sum^{N_y-1}_{v=0}\left(F_{u,v}\sin{C_{u,v}}\right)}{\sum^{N_x-1}_{u=0}\sum^{N_y-1}_{v=0}\left(F_{u,v}\cos{C_{u,v}}\right)}\right]}
    \end{equation}
    
    We can choose the correct solution by using $\nicefrac{\mathrm{d}^2\Delta E'}{\mathrm{d}\theta_{H'}^2} > 0$
    
    \begin{equation} 
    \cos{\theta_{H'}}\sum^{N_x-1}_{u=0}\sum^{N_y-1}_{v=0}F_{u,v}\cos{C_{u,v}} + \sin{\theta_{H'}}\sum^{N_x-1}_{u=0}\sum^{N_y-1}_{v=0}F_{u,v}\sin{C_{u,v}} > 0
    \end{equation}
    
\begin{figure}[htbp]
        \centering
        {\includegraphics[trim={0 0 0 0},width=0.7\linewidth,page=1]{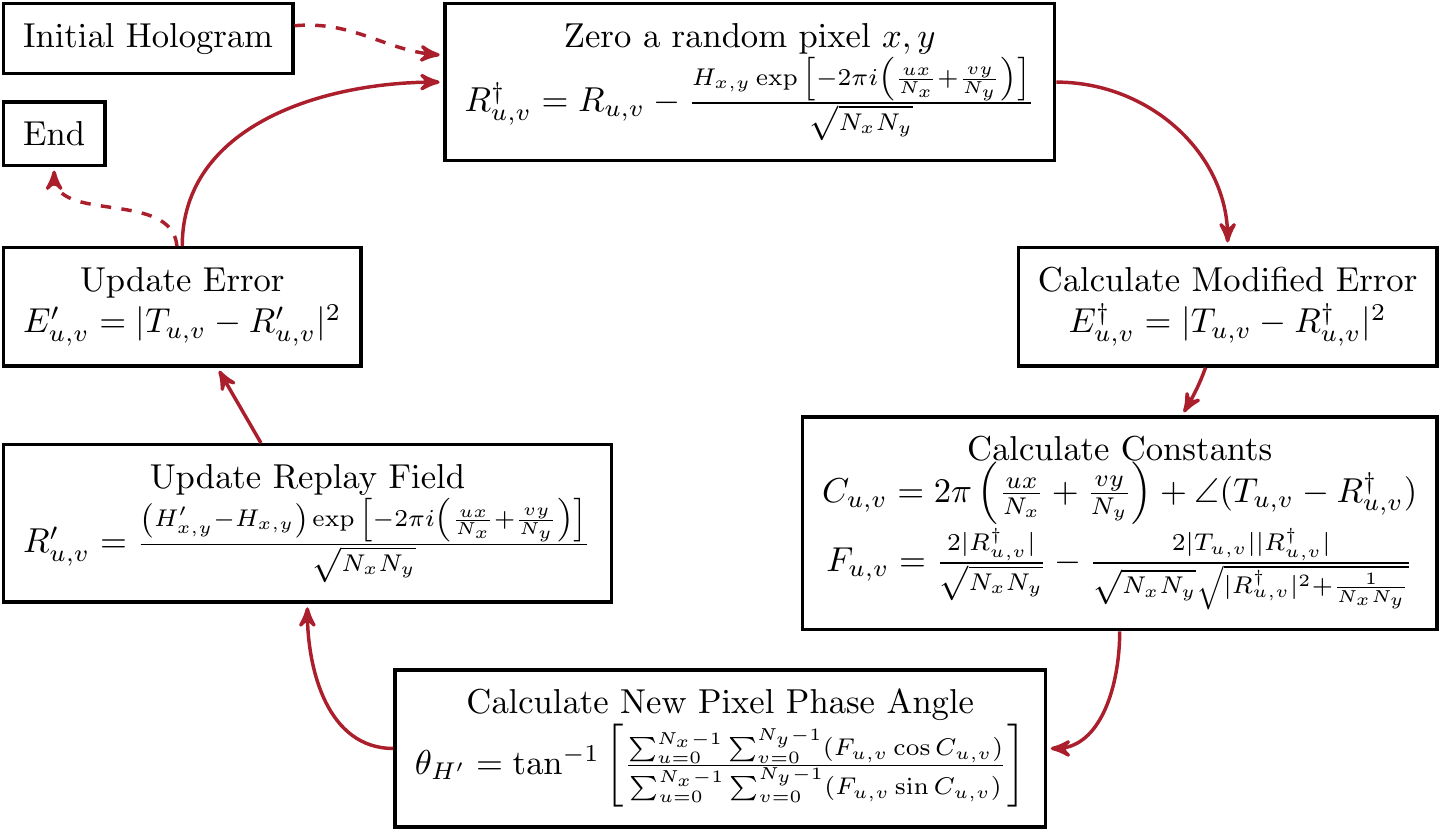}}
        \caption{Holographic Predictive Search for phase modulated, far-field holography with phase insensitive target} 
        \label{fig:alg_HPS_pi}
    \end{figure}

    This translates into the algorithm shown in Figure.~\ref{fig:alg_HPS_pi}
        
\begin{figure}[htbp]
        \centering
        {\includegraphics[trim={0 0 0 0},width=0.48\linewidth,page=1]{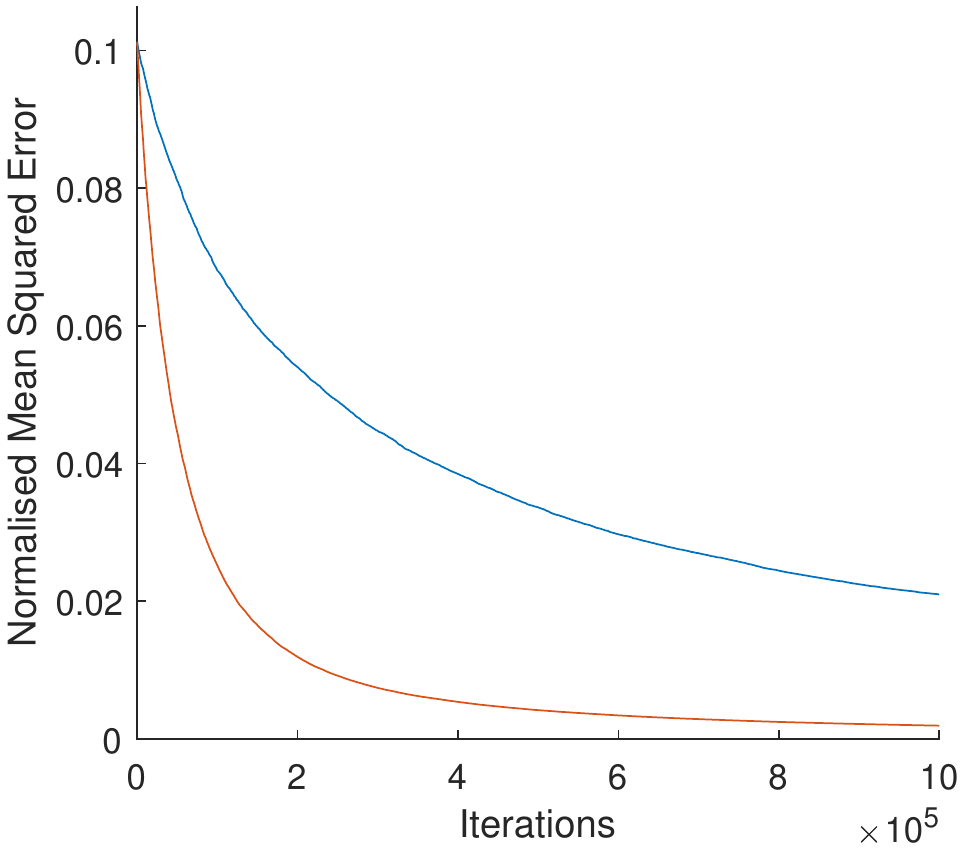}}
        {\includegraphics[trim={0 0 0 0},width=0.48\linewidth,page=1]{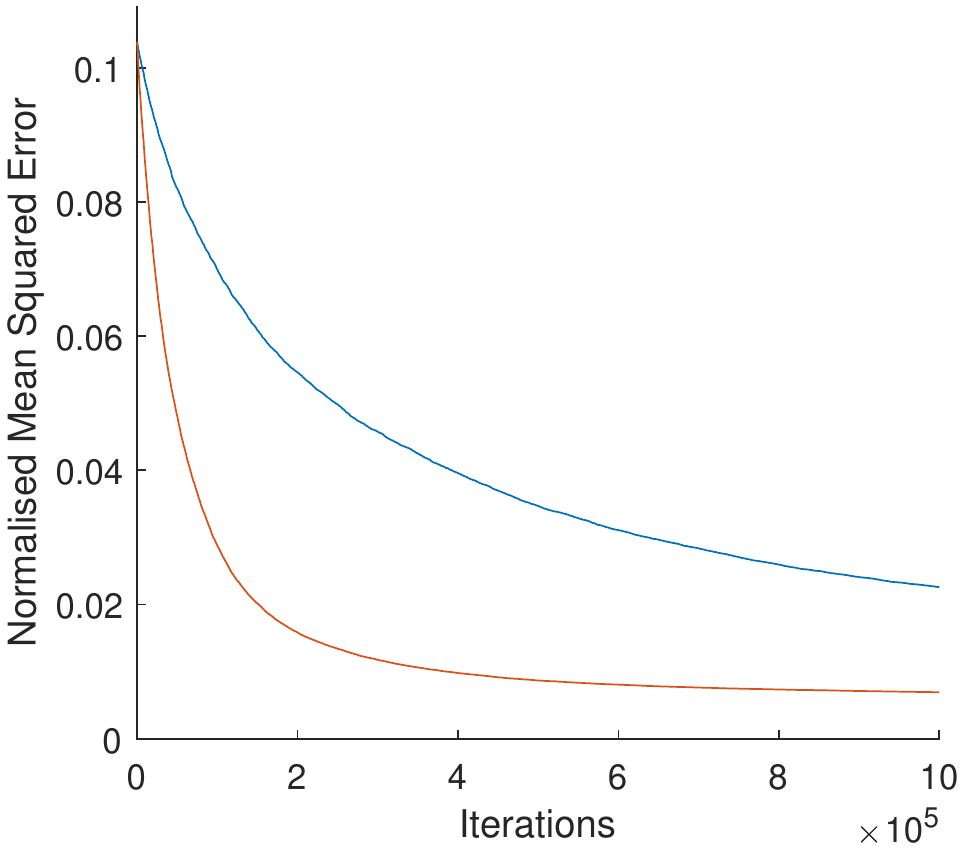}}
        \caption{Comparison of Direct Search (blue) against phase modulated, phase insensitive HPS (orange) for the $256\times256$ pixel \textit{Mandrill} test image being displayed on a $2^8$ (left) and $2^4$ (right) phase level spatial light modulator. Each trend line is taken as the mean of 5 independently seeded runs. }
        \label{fig:converge2}
    \end{figure}

    Tests of convergence for a $256\times256$ \textit{Mandrill} test image on a $2^8$ level phase SLM gives the performance graph as shown in Figure~\ref{fig:converge2}. The normalisation assumes a unit energy illumination to every SLM pixel and the target is scaled accordingly. 
    
    Comparison with the DS case presents a significant improvement in convergence speed but this is less marked than the phase sensitive case. This is not unexpected as an examination of Figure~\ref{fig:pixelChange_002} will show that the degree of variation per pixel is much lower in the PI case. i.e. the hologram initial hologram is much closer to the theoretical \textit{best} hologram than in the PS case. Nonetheless, the error at the end of $100,000$ iterations is $10\times$ lower for HPS than for DS.
    
    Figure~\ref{fig:converge2} also provides a comparison of the $2^8$ modulation level (left) vs $2^4$ modulation level (right) cases. This shows a similar performance improvement in both cases. We return to the effect of number of modulation levels later, but for now it should be noted that the primary difference is in an increased convergent error with other features remaining indistinguishable.
    
    \begin{figure}[htbp]
        \centering
        {\includegraphics[trim={0 0 0 0},width=1.0\linewidth,page=1]{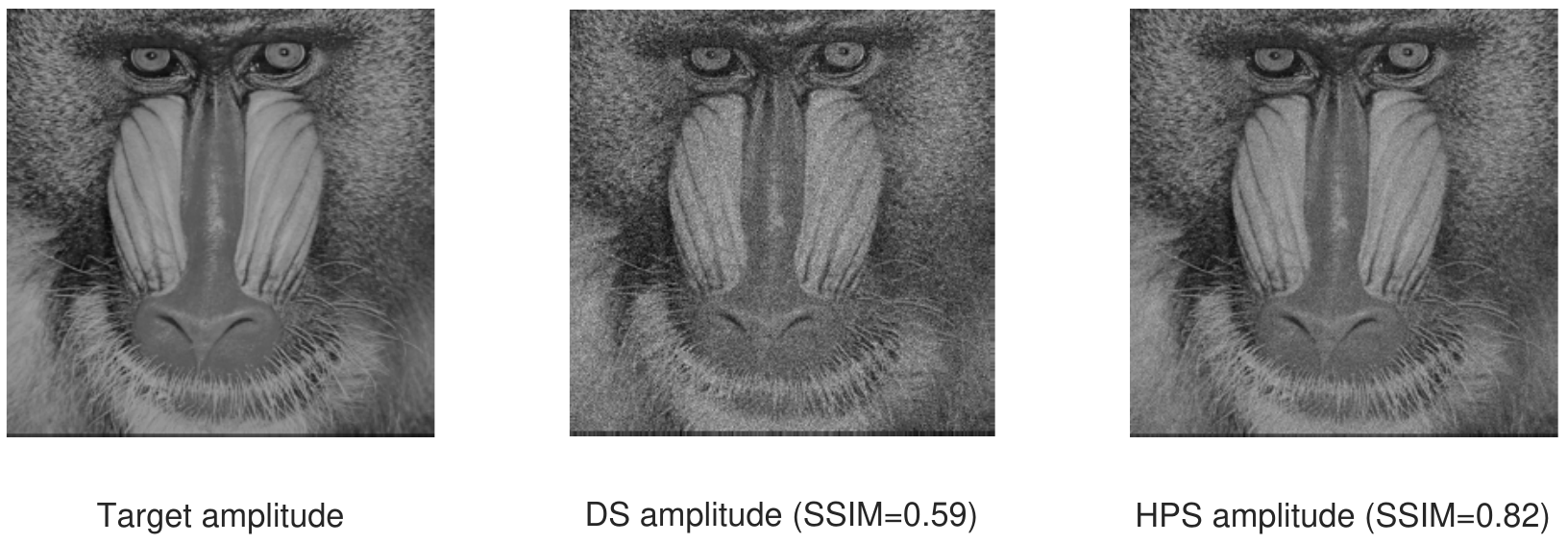}}
        \caption{Comparison of Direct Search (centre) against phase modulated, phase insensitive HPS (right) for $256\times256$ pixel \textit{Mandrill} test image (left) being displayed on a $2^8$ phase level spatial light modulator. Both algorithms were run for $1,000,000$ iterations.}
        \label{fig:converge2a}
    \end{figure}

    The algorithm reconstruction is shown in Figure~\ref{fig:converge2a} showing target image (left), DS (centre) and HPS (right). The SSIM values given are calculated with a dynamic range of $1.0$.
    
    Note also the use of the Taylor series expansion. An exact solution quickly devolves into a quartic polynomial which, while solvable, proves expensive computationally. Note also that it would be entirely possible to use more terms of the Taylor series for increased accuracy at the cost of performance. Our initial tests suggested that the use of only two terms was sufficient with error in predicted angle $\theta_{H'}$ never going above $1\%$. 
    
    Finally, we note that while HPS compares favourable with DS in this case, it is unlikely to offer benefits over Gerchberg-Saxton for high numbers of modulation levels \cite{gerchberg1972practical}. We return to this discussion later. 

    \subsection{Amplitude Modulated SLM, Phase Sensitive Replay Field} \label{as}

\begin{figure}[tb]
        \centering
        {\includegraphics[trim={0 0 0 0},width=0.7\linewidth,page=1]{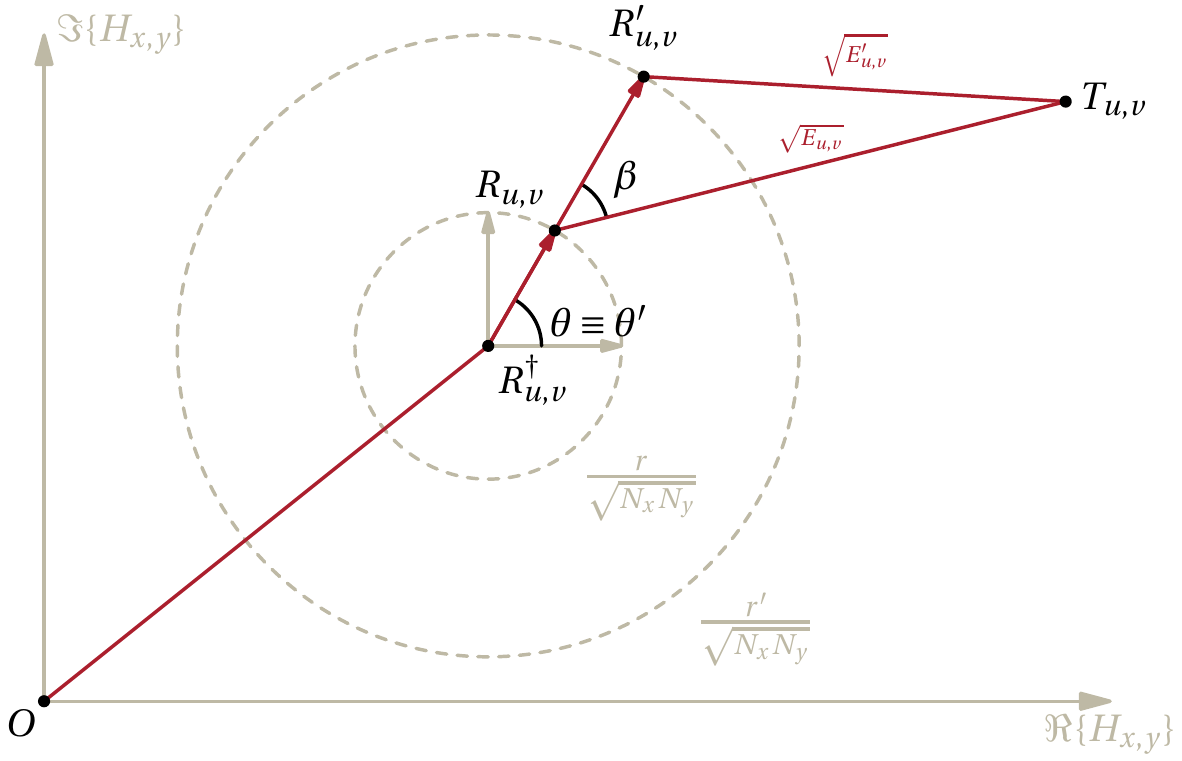}}
        \caption{Problem geometry in amplitude modulated, phase sensitive case.}
        \label{fig:predicted3}
    \end{figure}

    Similar to its phase modulated counterpart, amplitude modulated phase sensitive HPS has a problem geometry as shown in Figure~\ref{fig:predicted3}. Unlike in the phase modulated case, the new pixel phase angle $\theta_{H'}$ is equal to zero. This allows us to skip the step of zeroing a pixel and ignore $R^{\dagger}$ and $E^{\dagger}$. Instead we can work in terms of $\Delta r'=r'-r$ where $r$ and $r'$ are the old and new pixel magnitudes respectively.
    
    The initial error before modifying pixel $x,y$ is given as $E_{u,v}  = \abs{T_{u,v}-R_{u,v}}^2$ which is knowable at runtime. The new error $E'_{u,v}$ is given by
    
    \begin{align} 
    E'_{u,v} & = \abs{T_{u,v}-R'_{u,v}}^2 \nonumber \\
    & = \left[\abs{T_{u,v}-R_{u,v}}-\frac{\Delta r}{\sqrt{N_xN_y}}\cos{\beta}\right]^2 + \left[\frac{\Delta r}{\sqrt{N_xN_y}}\sin{\beta}\right]^2  \nonumber \\
    & = E_{u,v}
    + \frac{\Delta r^2}{N_xN_y}\cos^2{\beta}
    - 2\sqrt{E_{u,v}}\frac{\Delta r}{\sqrt{N_xN_y}}\cos{\beta}
    + \frac{\Delta r^2}{N_xN_y}\sin^2{\beta}
    \end{align}
    
    Remembering that $\cos^2{\beta}+\sin^2{\beta}=1$, the change in error for given $\frac{\Delta r}{\sqrt{N_xN_y}}$ is
    
    \begin{align} \label{eqn222}
    \Delta E'_{u,v} & = E'_{u,v} - E_{u,v} \nonumber \\
    & = \frac{\Delta r^2}{N_xN_y}
    - 2\sqrt{E_{u,v}}\frac{\Delta r}{\sqrt{N_xN_y}}\cos{\beta}
    \end{align}
    
    $\beta_{u,v}$ is given from $\theta_{u,v}$
    
    \begin{equation} 
    \beta_{u,v} = \theta_{u,v} - \angle(T_{u,v}-R_{u,v}) = -2\pi\left(\frac{ux}{N_x}+\frac{vy}{N_y}\right) - \angle(T_{u,v}-R_{u,v})
    \end{equation}
    
    where
    
    \begin{equation}
    \theta' = \theta = -2\pi\left(\frac{ux}{N_x}+\frac{vy}{N_y}\right)
    \end{equation}
    
    Summing 
    
    \begin{align}
    \Delta E' &= \sum^{N_x-1}_{u=0}\sum^{N_y-1}_{v=0} \Delta E'_{u,v} \nonumber \\
    & = \Delta r^2
    - \frac{2 \Delta r}{\sqrt{N_xN_y}}
    \sum^{N_x-1}_{u=0}\sum^{N_y-1}_{v=0}\sqrt{E_{u,v}}\cos{\beta} 
    \end{align}
    
    Taking $\nicefrac{\mathrm{d}\Delta E'}{\mathrm{d}\Delta r}=0$ to find the the value of $\Delta r$ where $\Delta E'$ is minimum
    
    \begin{equation} 
    \Delta r  = \frac{1}{\sqrt{N_xN_y}}
    \sum^{N_x-1}_{u=0}\sum^{N_y-1}_{v=0}\sqrt{E_{u,v}}\cos{\beta} 
    \end{equation}
    
\begin{figure}[htbp]
        \centering
        {\includegraphics[trim={0 0 0 0},width=0.7\linewidth,page=1]{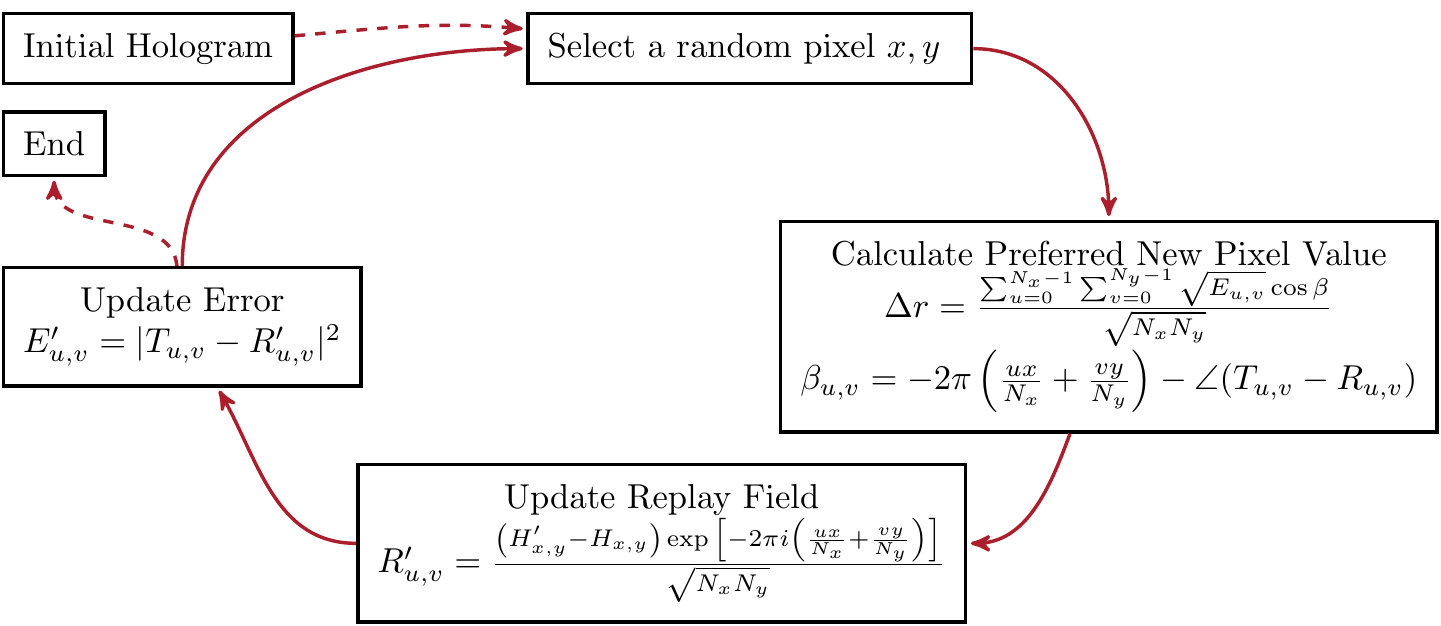}}
        \caption{Holographic Predictive Search for amplitude modulated, far-field holography with phase sensitive target.} 
        \label{fig:alg_HPS_as}
    \end{figure}

    The linear nature of this result means that we can cap $\Delta r$ within the constraints of the SLM. This is shown algorithmically in Figure~\ref{fig:alg_HPS_as}.
    
\begin{figure}[htbp]
        \centering
        {\includegraphics[trim={0 0 0 0},width=0.48\linewidth,page=1]{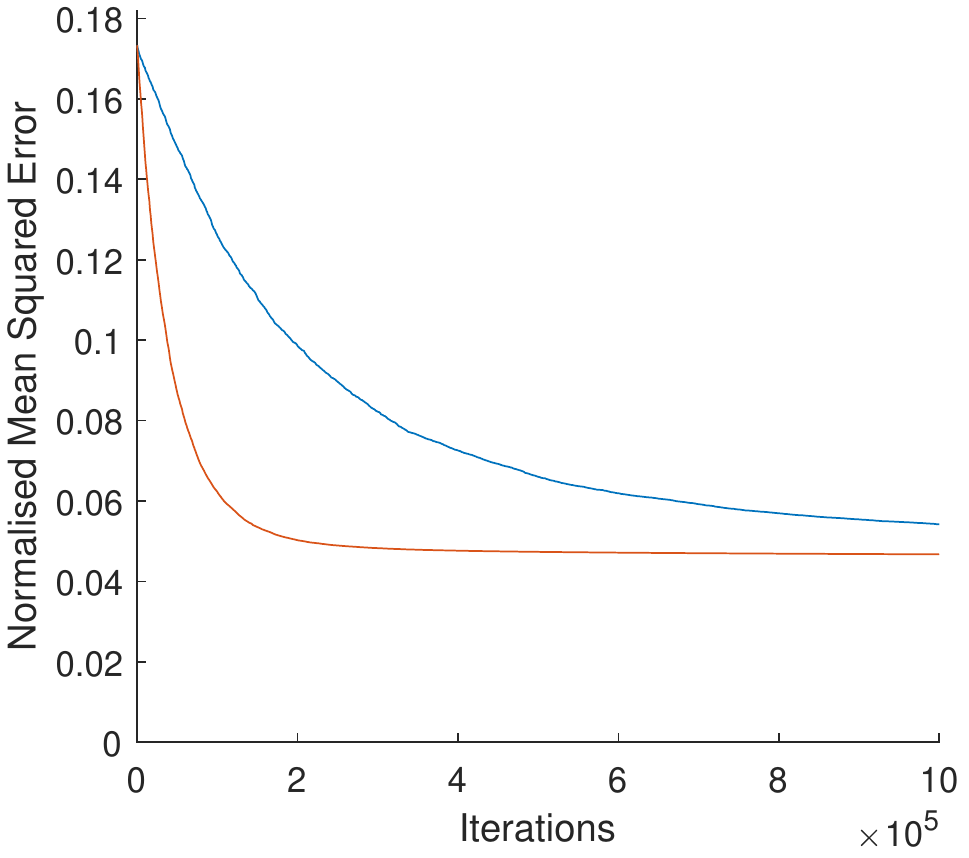}}
        {\includegraphics[trim={0 0 0 0},width=0.48\linewidth,page=1]{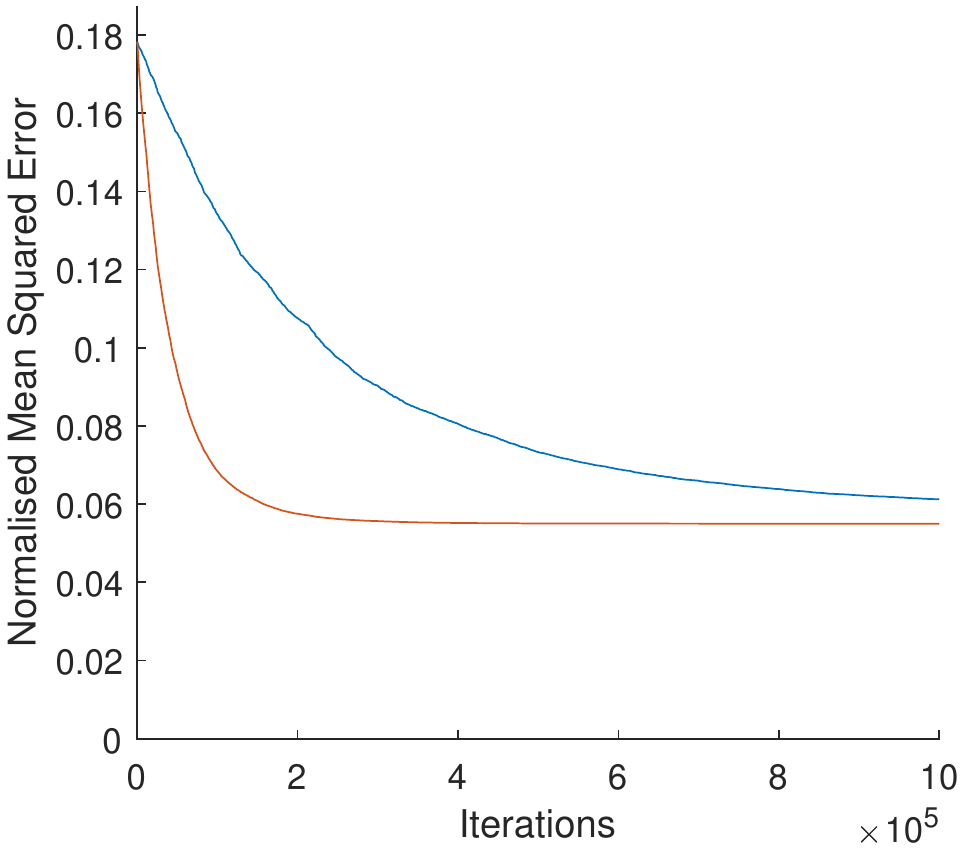}}
        \caption{Comparison of Direct Search (blue) against amplitude modulated, phase sensitive HPS (orange) for the $256\times256$ pixel \textit{Mandrill} test image being displayed on a $2^8$ (left) and $2^4$ (right) amplitude level spatial light modulator.  Each trend line is taken as the mean of 5 independently seeded runs.}
        \label{fig:converge3}
    \end{figure}

    This is modelled for a $256\times256$ \textit{Mandrill} test image with target phase given by the \textit{Peppers} test image. On a simulated $2^8$ level amplitude SLM this gives the performance graph as shown in Figure~\ref{fig:converge3}. This results in an approximately $2\times$ improvement in convergence speed to reach a given target error. 
    
\begin{figure}[htbp]
        \centering
        {\includegraphics[trim={0 0 0 0},width=1.0\linewidth,page=1]{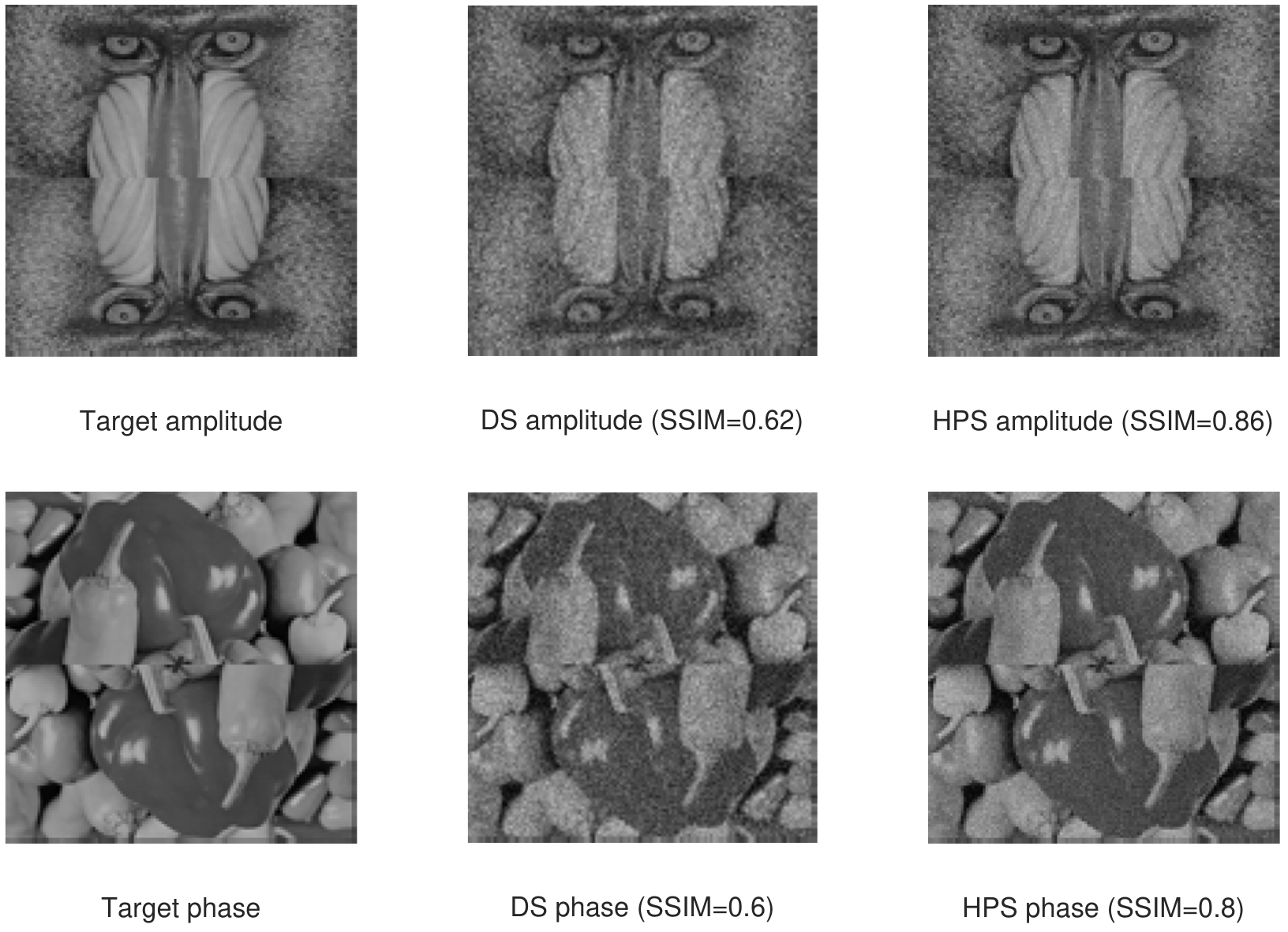}}
        \caption{Comparison of Direct Search (centre) against amplitude modulated, phase sensitive HPS (right) for $128\times 128$ pixel \textit{Mandrill} test image amplitudes (top left) and \textit{Peppers} test image phases (top left) being displayed on a $2^8$ amplitude level $256\times256$ pixel spatial light modulator. Both algorithms were run for $1,000,000$ iterations.}
        \label{fig:converge3a}
    \end{figure}

    The algorithm reconstruction is shown in Figure~\ref{fig:converge3a} showing target image (left), DS (centre) and HPS (right). The SSIM values given are calculated with a dynamic range of $1.0$.
    
    Very similar performance improvements were seen when used in place of a simulated annealing algorithm where HPS consistently outperformed SA approximately $2\times$ in terms of iterations required to reach a given target error. While this is interesting, it should be noted that phase sensitive amplitude holography is unlikely to be a common paradigm and this result is presented primarily for completeness. We return to this later.

    \subsection{Amplitude Modulated SLM, Phase Insensitive Replay Field} \label{ai}
       
\begin{figure}[tb]
        \centering
        {\includegraphics[trim={0 0 0 0},width=0.7\linewidth,page=1]{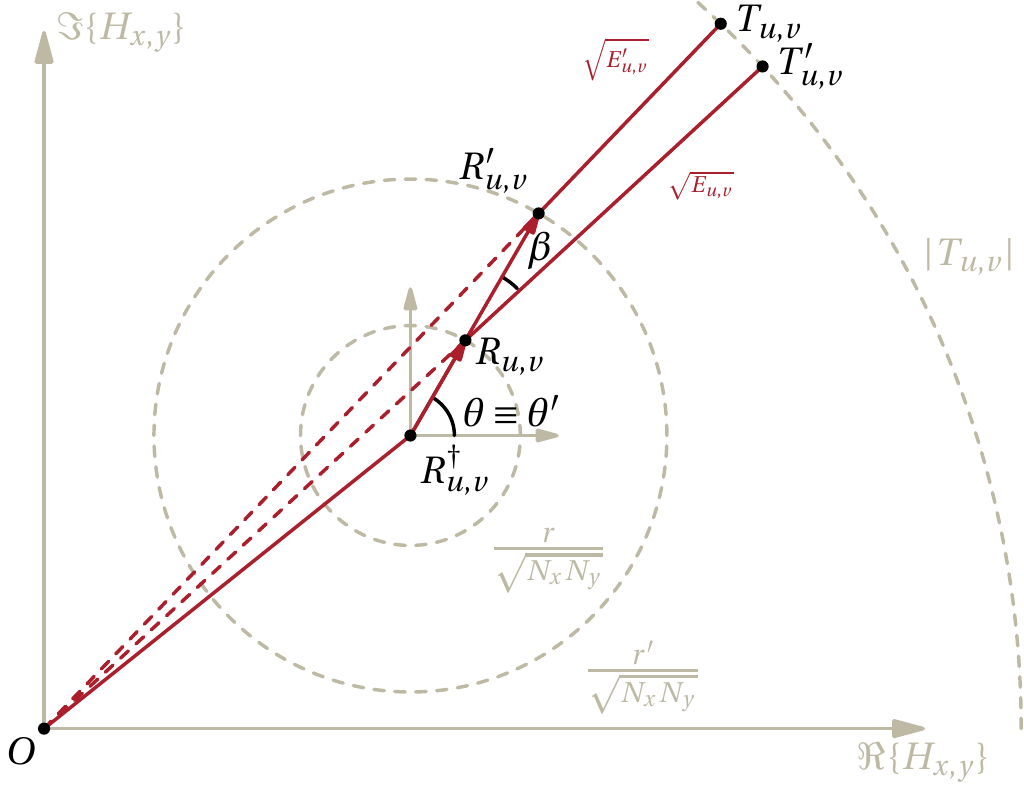}}
        \caption{Problem geometry in amplitude modulated, phase insensitive case.}
        \label{fig:predicted4}
    \end{figure}

    The phase insensitive amplitude modulated behaviour is similar to the phase insensitive phase modulated case. The problem geometry is shown in Figure~\ref{fig:predicted4}. Working again in terms of $\Delta r'=r'-r$ where $r$ and $r'$ are the old and new pixel magnitudes respectively.
    
    \begin{align}
        E'_{u,v} & = \left[\abs{T_{u,v}}-\abs{R'_{u,v}}\right]^2 \nonumber \\
        &= \left[\abs{T_{u,v}}-\sqrt{\abs{R_{u,v}}^2+\frac{\Delta r^2}{N_xN_y}+2\abs{R_{u,v}}\frac{\Delta r}{\sqrt{N_xN_y}}\cos{\beta}}\right]^2 \nonumber \\
        \Rightarrow \Delta E'_{u,v} & =  E'_{u,v} - E_{u,v} \nonumber \\
        &= \frac{\Delta r^2}{N_xN_y}+2\abs{R_{u,v}}\frac{\Delta r}{\sqrt{N_xN_y}}\cos{\beta}+2\abs{T_{u,v}}\abs{R_{u,v}}\nonumber \\
        &-2\abs{T_{u,v}}\abs{R_{u,v}}\sqrt{\abs{R_{u,v}}^2+\frac{\Delta r^2}{N_xN_y}+2\abs{R_{u,v}}\frac{\Delta r}{\sqrt{N_xN_y}}\cos{\beta}}
    \end{align}

    Unlike in Section~\ref{pi}, there is no easy Taylor substitution. Instead we assume that ${\Delta r}^2+2\abs{R_{u,v}}{\Delta r}\cos{\beta}$ is smaller than $\abs{R_{u,v}}^2$. This assumption can be seen to be valid for almost all non-zero target replay field values. For the \textit{Mandrill} test image this results in $>99.99\%$ of pixels being valid with this assumption. For different amplitude distributions this assumption becomes less valid. Fortunately for us, however, the system is insensitive to such pixels as they have near zero magnitude and this is further improved by the square relationship between intensity and amplitude. In the tests run here we found less that $0.02\%$ of pixels for which this gave greater than $1\%$ error in target value.
    
    If we allow this assumption we can write
    
    \begin{align}
    \Delta E'_{u,v} & =  E'_{u,v} - E_{u,v} \nonumber \\
    &= (1-\abs{T_{u,v}})(\frac{\Delta r^2}{N_xN_y}+2\abs{R_{u,v}}\frac{\Delta r}{\sqrt{N_xN_y}}\cos{\beta})
    \end{align}
    
    Where $\beta_{u,v}$ is again given by
    
    \begin{equation} 
    \beta_{u,v} = \theta_{u,v} - \angle(T_{u,v}-R_{u,v}) = -2\pi\left(\frac{ux}{N_x}+\frac{vy}{N_y}\right) - \angle(T_{u,v}-R_{u,v})
    \end{equation}
        
    Summing 
    
    \begin{align}
    \Delta E' &= \sum^{N_x-1}_{u=0}\sum^{N_y-1}_{v=0} \Delta E'_{u,v} \nonumber \\
    & = (1-\sum^{N_x-1}_{u=0}\sum^{N_y-1}_{v=0} T_{u,v})\Delta r^2
    + \frac{2 \Delta r}{\sqrt{N_xN_y}}
    \sum^{N_x-1}_{u=0}\sum^{N_y-1}_{v=0}(1-T_{u,v})\abs{R_{u,v}}\cos{\beta}
    \end{align}
    
    Taking $\nicefrac{\mathrm{d}\Delta E'}{\mathrm{d}\Delta r}=0$ to find the the value of $\Delta r$ where $\Delta E'$ is minimum
    
    \begin{equation} 
    \Delta r  = \frac{1}{\sqrt{N_xN_y}}
    \sum^{N_x-1}_{u=0}\sum^{N_y-1}_{v=0}(1-T_{u,v})\abs{R_{u,v}}\cos{\beta}
    \end{equation}
    
\begin{figure}[htbp]
        \centering
        {\includegraphics[trim={0 0 0 0},width=0.7\linewidth,page=1]{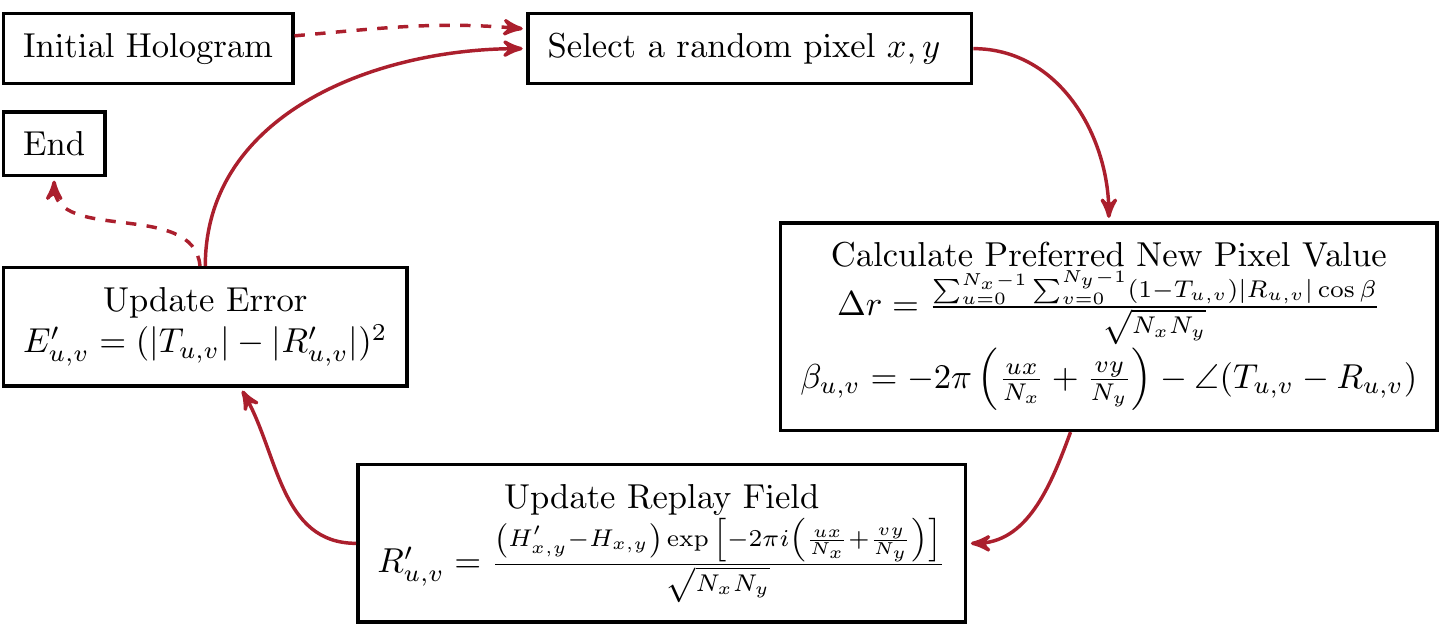}}
        \caption{Holographic Predictive Search for amplitude modulated, far-field holography with phase insensitive target.} 
        \label{fig:alg_HPS_ai}
    \end{figure}

    The linear nature of this result means that we can cap $\Delta r$ within the constraints of the SLM. This translates into the algorithm shown in Figure.~\ref{fig:alg_HPS_ai}
    
\begin{figure}[htbp]
        \centering
        {\includegraphics[trim={0 0 0 0},width=0.48\linewidth,page=1]{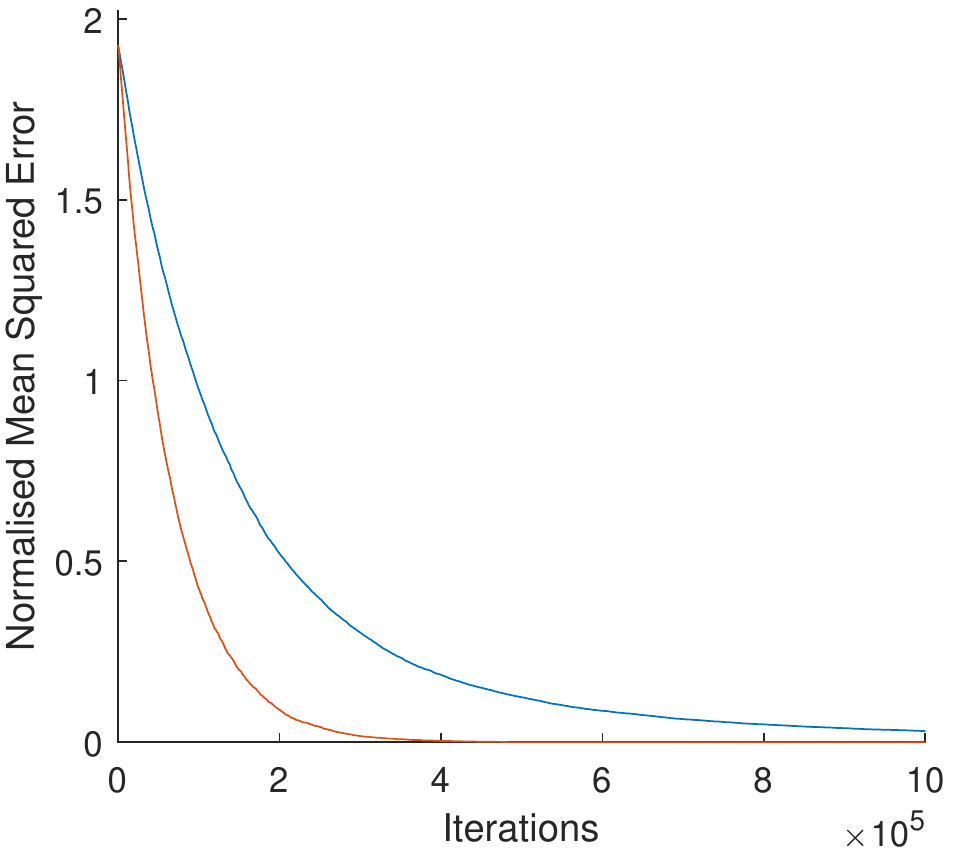}}
        {\includegraphics[trim={0 0 0 0},width=0.48\linewidth,page=1]{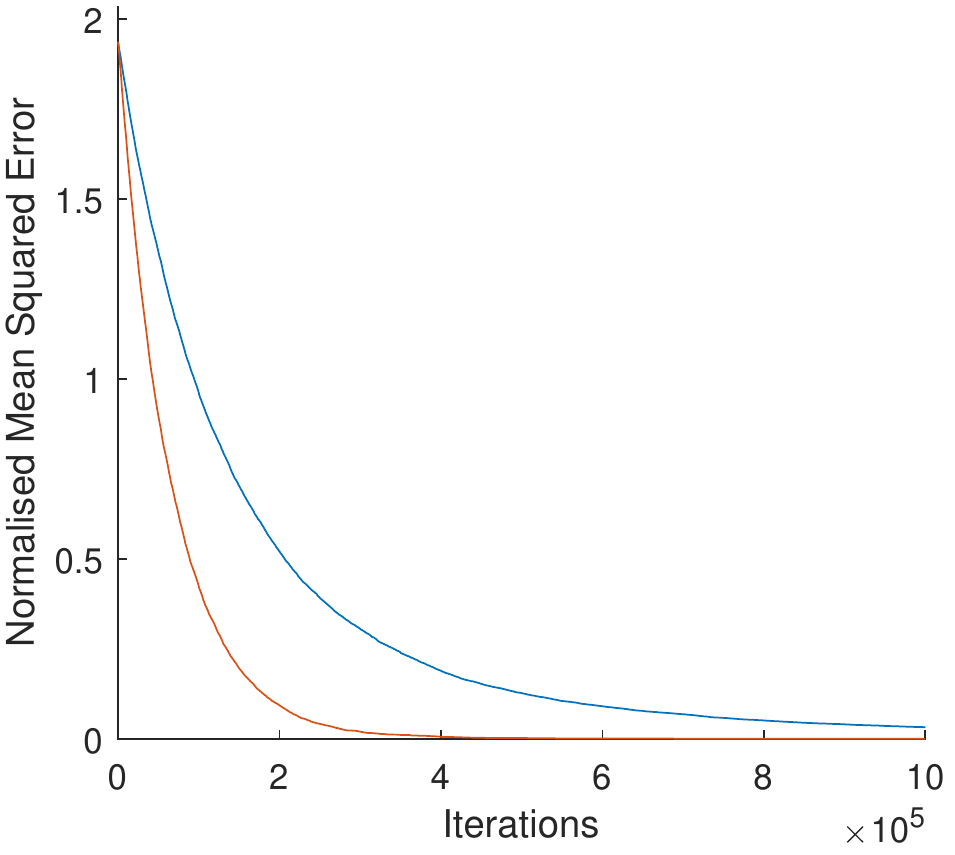}}
        \caption{Comparison of Direct Search (blue) against amplitude modulated, phase insensitive HPS (orange) for the $256\times256$ pixel \textit{Mandrill} test image being displayed on a $2^8$ (left) and $2^4$ (right) amplitude level spatial light modulator. Each trend line is taken as the mean of 5 independently seeded runs.}
        \label{fig:converge4}
    \end{figure}

    Tests of convergence for a $256\times256$ \textit{Mandrill} test image on a $2^8$ level amplitude SLM gives the performance graph as shown in Figure~\ref{fig:converge4}. This again results in an approximately $2\times$ improvement in convergence iterations to reach a given target error. 
  
\begin{figure}[htbp]
        \centering
        {\includegraphics[trim={0 0 0 0},width=1.0\linewidth,page=1]{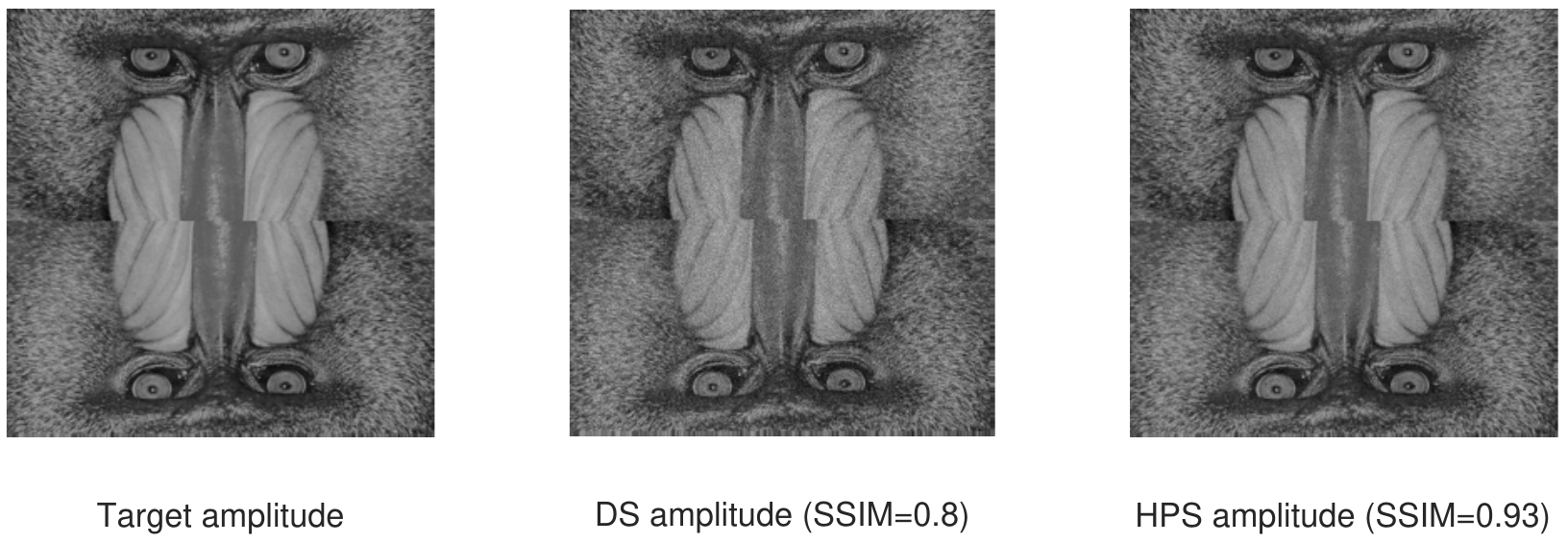}}
        \caption{Comparison of Direct Search (centre) against amplitude modulated, phase insensitive HPS (right) for $256\times256$ pixel \textit{Mandrill} test image (left) being displayed on a $2^8$ amplitude level spatial light modulator. Both algorithms were run for $1,000,000$ iterations.}
        \label{fig:converge4a}
    \end{figure}

    The algorithm reconstruction is shown in Figure~\ref{fig:converge4a} showing target image (left), DS (centre) and HPS (right). The SSIM values given are calculated with a dynamic range of $1.0$.
    
    \section{Fresnel Domain}

    Fortunately, the Fresnel transform variants of HPS turn out to have a very similar form to their Fraunhofer counterparts. The only distinction is the addition of the quadratic phase term to $C_{u,v}$ and $\beta_{u,v}$
    
    \begin{align} 
        C_{u,v} &= 2\pi\left(\frac{ux}{N_x}+\frac{vy}{N_y}\right) + \angle{R^{\dagger}_{u,v}} + e^{\frac{i \pi}{\lambda z}(x^2 + y^2)}  \nonumber \\
    \beta_{u,v} &= 2\pi\left(\frac{ux}{N_x}+\frac{vy}{N_y}\right) - \angle(T_{u,v}-R_{u,v}) + e^{\frac{i \pi}{\lambda z}(x^2 + y^2)}
    \end{align}
    
    where $z$ is the perpendicular separation between diffraction field and replay field and $\lambda$ is the illumination wavelength.

    \begin{table}[tbp]
        \centering
        \caption{Mathematical relationships for different HPS variants}
        \resizebox{\textwidth}{!}{%
            \begin{tabular}{c c}
                \toprule
                Phase Modulated SLM, Phase Sensitive Replay Field, Fraunhofer Domain              & Phase Modulated SLM, Phase Sensitive Replay Field, Fresnel Domain \\
                \midrule
                $\begin{aligned}[t] 
                C_{u,v} &= 2\pi\left(\frac{ux}{N_x}+\frac{vy}{N_y}\right) + \angle{R^{\dagger}_{u,v}} \nonumber \\
                \theta_{H'}&=\tan^{-1}{\left[\frac{\sum^{N_x-1}_{u=0}\sum^{N_y-1}_{v=0}\left(\sqrt{E^{\dagger}_{u,v}}\cos{C_{u,v}}\right)}{\sum^{N_x-1}_{u=0}\sum^{N_y-1}_{v=0}\left(\sqrt{E^{\dagger}_{u,v}}\sin{C_{u,v}}\right)}\right]}
                \end{aligned}$
                &
                $\begin{aligned}[t] 
                C_{u,v} &= 2\pi\left(\frac{ux}{N_x}+\frac{vy}{N_y}\right) + \angle{R^{\dagger}_{u,v}} + e^{\frac{i \pi}{\lambda z}(x^2 + y^2)} \nonumber \\
                \theta_{H'}&=\tan^{-1}{\left[\frac{\sum^{N_x-1}_{u=0}\sum^{N_y-1}_{v=0}\left(\sqrt{E^{\dagger}_{u,v}}\cos{C_{u,v}}\right)}{\sum^{N_x-1}_{u=0}\sum^{N_y-1}_{v=0}\left(\sqrt{E^{\dagger}_{u,v}}\sin{C_{u,v}}\right)}\right]}
                \end{aligned}$ \\
                \bottomrule
                Phase Modulated SLM, Phase Insensitive Replay Field, Fraunhofer Domain              & Phase Modulated SLM, Phase Insensitive Replay Field, Fresnel Domain \\
                \midrule
                $\begin{aligned}[t] 
                C_{u,v} &= 2\pi\left(\frac{ux}{N_x}+\frac{vy}{N_y}\right) + \angle{R^{\dagger}_{u,v}} \nonumber \\
                F_{u,v} &= \frac{2\abs{R^{\dagger}_{u,v}}}{\sqrt{N_xN_y}}-\frac{2\abs{T_{u,v}}\abs{R^{\dagger}_{u,v}}}{\sqrt{N_xN_y}\sqrt{\abs{R^{\dagger}_{u,v}}^2+\frac{1}{N_xN_y}}} \nonumber \\
                \theta_{H'}&=\tan^{-1}{\left[\frac{\sum^{N_x-1}_{u=0}\sum^{N_y-1}_{v=0}\left(F_{u,v}\cos{C_{u,v}}\right)}{\sum^{N_x-1}_{u=0}\sum^{N_y-1}_{v=0}\left(F_{u,v}\sin{C_{u,v}}\right)}\right]}
                \end{aligned} $&
                $\begin{aligned}[t] 
                C_{u,v} &= 2\pi\left(\frac{ux}{N_x}+\frac{vy}{N_y}\right) + \angle{R^{\dagger}_{u,v}} + e^{\frac{i \pi}{\lambda z}(x^2 + y^2)} \nonumber \\
                F_{u,v} &= \frac{2\abs{R^{\dagger}_{u,v}}}{\sqrt{N_xN_y}}-\frac{2\abs{T_{u,v}}\abs{R^{\dagger}_{u,v}}}{\sqrt{N_xN_y}\sqrt{\abs{R^{\dagger}_{u,v}}^2+\frac{1}{N_xN_y}}} \nonumber \\
                \theta_{H'}&=\tan^{-1}{\left[\frac{\sum^{N_x-1}_{u=0}\sum^{N_y-1}_{v=0}\left(F_{u,v}\cos{C_{u,v}}\right)}{\sum^{N_x-1}_{u=0}\sum^{N_y-1}_{v=0}\left(F_{u,v}\sin{C_{u,v}}\right)}\right]}
                \end{aligned} $ \\
                \bottomrule
                Amplitude Modulated SLM, Phase Sensitive Replay Field, Fraunhofer Domain              & Amplitude Modulated SLM, Phase Sensitive Replay Field, Fresnel Domain \\
                \midrule
                $\begin{aligned}[t] 
                \beta_{u,v} &  = 2\pi\left(\frac{ux}{N_x}+\frac{vy}{N_y}\right) - \angle(T_{u,v}-R_{u,v}) \nonumber \\
                \Delta r & = \frac{1}{\sqrt{N_xN_y}}
                \sum^{N_x-1}_{u=0}\sum^{N_y-1}_{v=0}\sqrt{E_{u,v}}\cos{\beta} 
                \end{aligned} $&
                $\begin{aligned}[t] 
                \beta_{u,v} &  = 2\pi\left(\frac{ux}{N_x}+\frac{vy}{N_y}\right) - \angle(T_{u,v}-R_{u,v}) + e^{\frac{i \pi}{\lambda z}(x^2 + y^2)} \nonumber \\
                \Delta r & = \frac{1}{\sqrt{N_xN_y}}
                \sum^{N_x-1}_{u=0}\sum^{N_y-1}_{v=0}\sqrt{E_{u,v}}\cos{\beta} 
                \end{aligned} $ \\
                \bottomrule
                Amplitude Modulated SLM, Phase Insensitive Replay Field, Fraunhoferr Domain              & Amplitude Modulated SLM, Phase Insensitive Replay Field, Fresnel Domain \\
                \midrule
                $\begin{aligned}[t] 
                \beta_{u,v} &  = 2\pi\left(\frac{ux}{N_x}+\frac{vy}{N_y}\right) - \angle(T_{u,v}-R_{u,v}) \nonumber \\
                \Delta r & = \frac{1}{\sqrt{N_xN_y}}
                \sum^{N_x-1}_{u=0}\sum^{N_y-1}_{v=0}(1-T_{u,v})\abs{R_{u,v}}\cos{\beta}
                \end{aligned} $&
                $\begin{aligned}[t] 
                \beta_{u,v} &  = 2\pi\left(\frac{ux}{N_x}+\frac{vy}{N_y}\right) - \angle(T_{u,v}-R_{u,v}) + e^{\frac{i \pi}{\lambda z}(x^2 + y^2)} \nonumber \\
                \Delta r & = \frac{1}{\sqrt{N_xN_y}}
                \sum^{N_x-1}_{u=0}\sum^{N_y-1}_{v=0}(1-T_{u,v})\abs{R_{u,v}}\cos{\beta}
                \end{aligned} $ \\
                \bottomrule
            \end{tabular}
        }
        \label{table1}
    \end{table}
       
    \section{Summary, Discussion and Recommendations}
            
    The required relationships for the different HPS variants are summarised in Table~\ref{table1}. While we have briefly presented the relationships behind HPS, there are a significant number of points that should be discussed.
                
    Firstly, competitor algorithm families should be considered. HSAs such as DS can be used to generate some of the best quality holograms, albeit at the expense of slower generation times and HPS uniformly out-performed DS for every case discussed in this paper in both speed and convergent quality. 
    
    Iterative algorithms - for example Gerchberg-Saxton~\cite{gerchberg1972practical} - are available for many relatively smooth systems and should be expected to be significantly faster than HSAs including HPS. This comes at the expense of final image quality where HPS is expected to still give the best performance. For more-discontinuous systems with lower numbers of modulation levels, iterative algorithms can fail to converge and HSAs become a suitable alternative. In this case HPS may be expected to offer better performance in speed as well as quality.
    
    Secondly, the computational performance of HPS should be considered. In our previous work we found that HPS required approximately $70-80\%$ more time per iteration though this dropped to as little as $10\%$ in the case of larger images where computation was memory bound. This increased iteration time should be taken into account when selecting an appropriate algorithm.
    
    Thirdly, this study has been purely mathematical in nature. Account has not been taken for real-world imperfections such as lens aberration, non-flatness or speckle. The authors suggest that these effects are likely to affect all HSAs similarly but recommend further study of the sensitivity of different algorithms to real-world errors.
    
    The fourth point is that the image quality metric used here is MSE. SSIM has seen increased use in recent years as it more closely corresponds to the human eye behaviour. While recent authors have argued for a more close relationship between MSE and SSIM than thought previously \cite{Dosselmann2011} the authors acknowledge that this is a weakness in the HPS method and propose further investigation.
    
    The final point regards the complexity of the HPS method. While in certain situations HPS offers significant performance improvements over rival techniques, the increased complexity and reduced generality will require a greater level of expertise than alternative techniques. 
    
    \section{Conclusion}
    
    This work has presented seven new variants on the Holographic Predictive Search (HPS) algorithm which are summarised in Table~\ref{table1}. By using this, prescient search techniques can be used for a wide range of optical systems in both the far- and mid-field. Different modulation schemes and replay field constraints have all been discussed. 
    
    When compared to direct search and simulated annealing algorithms, HPS has been shown to be over $10\times$ faster than its competitors in specific cases at the expense of increased complexity and reduced flexibility. HPS also offers the best convergent error quality. Variants on the HPS algorithm have been presented for a range of optical configurations and the relative advantages and disadvantages presented. 
    
    \section*{Funding}
    
    The authors would like to thank the Engineering and Physical Sciences Research Council (EP/L016567/1 and EP/L015455/1) for financial support during the period of this research.
    
    \section*{Disclosures}
    
    The authors declare no conflicts of interest.
                
    \section*{References}
        
    \bibliography{references}
    
\end{document}